\documentclass{aa}  

\usepackage{natbib}
\usepackage{graphicx}
\usepackage{amsmath}
\usepackage{mhchem}
\usepackage{xcolor}
\usepackage{ulem} 
\usepackage{caption}
\usepackage{float}
\usepackage[colorlinks=true, allcolors=blue]{hyperref}
\usepackage{soul}

\usepackage{makecell}

\usepackage{txfonts}
\begin{document} 

    \title{Ice chemistry that can be unveiled with the JWST: SynthIceSpec, a synthetic spectrum generator to test spectral limits}
   \subtitle{Solid CO$_2$ as a dust thermometer and solid CH$_3$CN detectability in cold cores}

   \author{A. Taillard\inst{1,2}, P. Gratier\inst{1}, J. A. Noble\inst{3}, E. Dartois\inst{4}, A. C. A. Boogert\inst{5}, J. Keane\inst{5}, A. Clément\inst{1}, A. Maiolo\inst{6,7}, A. Fuente \inst{2}, V. Wakelam\inst{1}
          } 

  \institute{Laboratoire d'Astrophysique de Bordeaux (LAB), Univ. Bordeaux, CNRS, B18N, allée Geoffroy Saint-Hilaire, 33615 Pessac, France
              \email{angele.taillard@cab.inta-csic.es}
         \and
        Centro de Astrobiología (CAB), CSIC-INTA, Ctra. de Ajalvir, km 4, Torrejón de Ardoz, 28850 Madrid, Spain             
         \and 
             Physique des Interactions Ioniques et Moléculaires, CNRS, Aix Marseille Univ., 13397 Marseille, France
         \and
             Institut des sciences Moléculaires d'Orsay, CNRS, Université Paris-Saclay, Bât 520, Rue André Rivière, 91405 Orsay, France
         \and
            Institute for Astronomy, 2680 Woodlawn Drive, Honolulu, HI 96822-1897, USA
         \and
            CEA-CESTA, Le Barp 33114,  France
         \and
            CELIA, University of Bordeaux-CNRS-CEA, Talence 33405, France
             }
   \date{To be submitted}

 
  \abstract 
   {
    
    As the \textit{James Webb Space Telescope} (JWST) pursues its observing journey, several thousands of icy-grain spectra are expected to be measured and analysed. The inventory of ices in particular, via the observations of background sources, is accessible for hundreds of lines of sight (LOSs) per molecular-cloud region, opening the possibility to add strong constraints on the solid phase chemistry in a vast domain of cloud densities.
    SynthIceSpec is a synthetic infrared (IR) spectrum generator that has been designed as a tool to support observing proposals and to test the outcome of chemical models. It is based on laboratory measurements of pure and mixed ices, where each vibrational component is fitted by a sum of Gaussian profiles. Given an initial ice chemical composition (either set by the user or the outputs of a chemical model), a full JWST spectrum is generated, to which the contribution of silicates; continuum, stellar photospheric absorption bands; and extinction law can be added. For the continuum, stellar photospheric models for a wide range of spectral types can be selected by the program, or, Spectral Energy Distribution (SEDs). We present a few use cases of SynthIceSpec: we probed the impact of dust temperature on CO$_2$ ice formation using IR data and gas-grain modelling. 
    Next, we used SynthIceSpec to explore the detectability of the main feature of CH$_3$CN at 4.45 $\mu$m in a cold core environment with the JWST, which was previously tentatively detected in YSOs. The detection thresholds we derive are reasonably low and observable, but identification is directly impacted by the photosphere absorptions that can greatly hinder identification. For some photostellar types, it could remain feasible. 
    Coupled with the Estimated Time Calculator of the Space Telescope Science Institute, SynthIceSpec can be used to find the optimum observational setup for new observations.
   }


   \titlerunning{SynthIceSpec, a synthetic spectrum generator to test spectral limits}
   \authorrunning{Taillard et al.}

   \maketitle
%

\section{Introduction}\label{intro}

Interstellar dust grains play a pivotal role in the formation of complex organic molecules (COMs) in the interstellar medium (ISM). They allow the condensation of gas-phase molecules, radicals, and atoms on their surface and lead to the formation of ice mantles in high-density environments (n$\rm_H$ $\gtrsim$ 10$^4$ cm$^{-3}$) with low temperatures (T $\lesssim$ 90 K), low stellar UV irradiation (G$_0$ $\lesssim$ 0.07), and high visual extinction (A$_V$ $\gtrsim$ 1.6 mag) \citep{Boogert_2015}.
Ices are composed of various volatile species where surface and bulk reactions occur and provide new species to the ISM (H$_2$O, CO, CO$_2$, CH$_3$OH, NH$_3$). Solid-phase chemistry allows the formation of COMs in dense clouds, including species forming only through surface and bulk chemistry and released into the gas phase \citep{Tielens_1982,Herbst_2009A_COMs}. The energetic and thermal processing of ices allow an increase in the complexity of the molecular composition in cold environments, where gas chemistry is limited. These processes also change the physical nature of molecular matrices, such as the crystallisation of H$_2$O \citep{Smith_1989} or CO$_2$ segregation \citep{Dartois_1998,Gerakines_1999}.

The observation of ices in the ISM is done almost exclusively via the near- and mid-infrared (NIR and MIR), targeting the vibrational transitions of functional groups. Each species can present different vibrational modes, which are mainly affected by the geometry of the molecule, the lattice configuration, and the change in electric dipole moment of the species. The main different types of vibrations are stretching, bending, and hindered rotation. 
In the context of interstellar-ice observations, infrared (IR) spectroscopy is the best analytical method to use when identifying solid-phase species and relying mainly on laboratory characterisation \citep[e.g. for instance][]{1993ApJS...86..713H,1995A&A...296..810G,1999A&A...343..966S}. The presence of different kinds of molecules in the ice matrix influences the vibrational modes of their neighbours, affecting the profile and the deduced solid-phase composition. 
The mixing measured in laboratory experiments is key to understanding the changes in the band's shape, position, and strength, which could hinder the possible identification of other species \citep{1993ApJS...86..713H}. 
Since the photons received in the instruments probe the entirety of the matter present along a line of sight (LOS), the observed spectrum is the result of absorption and scattering, which can also depend on grain size and shape. The interpretation of data must be done by taking into account that, in the simplest observation configuration, the final spectrum obtained is all the different kinds of material (ice, bulk, bare grains) probed between the observer and the background source. It leads to a difficulty in the interpretation of extinction spectra, and with the launch of the \textit{James Webb Space Telescope} (JWST), the number of LOSs observed is increasing rapidly. Thankfully, the laboratory experimentalists are providing plethora of spectra to analyse and disentangle the observations.

The searches for ice in the ISM started with the study of the H$_2$O band at 3.0 $\mu$m in the diffuse medium \citep{Danielson_1965, Knacke_1969,Gillett_Forrest_1973}.
Later, CO  was detected towards the protostar W33A \citep{Soifer_1979}. The detection of CO$_2$ was more difficult because its main absorption feature (stretching mode) at 4.27 $\mu$m falls in the telluric absorption window of the Earth's atmosphere. It was with the launch of the space-based InfraRed Astronomical Satellite (IRAS) that the bending mode of CO$_2$ at 15.2 $\mu$m was detected \citep{DHendecourt_1989, Whittet_1991}. The first detection of its stretching mode at 4.27 $\mu$m had to wait until  the launch of the Infrared Space Observatory (ISO) \citep{deGraauw_1996,DHendecourt_1996,Guertler_1996}. 
Since then, an increasing number of molecules have been detected, such as CH$_4$ \citep{Boogert_1996}, NH$_3$ \citep{Gibb_2000}, HCOOH, and H$_2$CO \citep{Keane_2001}. The simplest of the COMs, CH$_3$OH, was first detected from ground-based telescopes \citep{Chiar_1996} and widely observed with ISO and the \textit{Spitzer} space telescopes \citep{dartois_1997,Pontoppidan_2003,Boogert_2011}. 
The JWST has been adding upper-limits to some molecules \citep[SO$_2$, NO$_2,$ and COMs,][]{Mcclure_2023,Rocha_2024,Rocha2025}.
A big step in the characterisation of ice structure has been reached with the solid detection of the dangling OH ice feature at 2.7 $\mu$m \citep{Noble_2024}.
We are hopeful that more detections will be solidly confirmed and upper-limits added on more species with upcoming JWST observations. 

The search for new molecules has to be guided by the experimental work and theoretical calculations, providing decisive data to identify absorption bands.
Chemical models can also provide valuable information on which molecule to look for and in which environments. The work done in these three fields has to be coordinated together to investigate the thousands of LOSs that will be observed by the JWST. 

In this paper, we present our synthetic ice spectra generator, namely SynthIceSpec, aiming to produce vibrational spectra with more than 25 species listed in our database. Our tool is designed to help the preparation and interpretation of JWST observations and predict the detection of new potential species, with the possibility to add background stars and Young Stellar Objects (YSOs) continua to the absorption spectrum.
In Section~\ref{sis-program}, we present the code, the approximations, and the various features added so far. We then provide use cases of what can be done with SynthIceSpec in Section~\ref{use_case}; we first showcase chemical model predictions, then how it can be used for comparison with observations, and finally an exploration of the detectability of species yet to be found in ices.

\section{Synthetic ice spectrum generator}\label{sis-program}

In this section, we present our synthetic ice spectrum generator, namely SynthIceSpec. The tool is designed to produce ice spectra from laboratory data from an input ice composition (through column densities), to predict possible observations and to help in the interpretation of data with providing the position, band strength, and bandwidth of more than 25 different species at the moment. The code takes the column densities of the considered species as input and can produce a simple vibrational spectrum. Instrumental noise can be added, while stellar photospheres, extinction, and modified black-body continuum can be taken into account. The output is an ascii file, providing the flux (in milliJansky) as a function of the wavelength (in microns). SynthIceSpec is available publicly on a github repository\footnote{\url{https://forge.oasu.u-bordeaux.fr/LAB/astrochem-tools/synthicespec.git}} and is open-source.

\subsection{Band parameters and column-density computation}

Ices are traced by their vibrational transitions in the NIR, MIR and far-IR. In the case of the JWST, the wavelengths observed are ranging from 0.6 $\mu$m to 28.3 $\mu$m combining both NIRSpec and MIRI. 
To reproduce the ice composition, we make the simple assumption that the shape of each functional group can be modelled by a Gaussian profile or a series of Gaussian profiles, characterised by their width, band strength, and peak position. The band intensity can then be computed by multiplying the assumed column density by the corresponding band strength and divided by the full width at half maximum (FWHM). 
A Gaussian profile, linking the optical depth, $\tau,$ of an absorption band to the column density, N, can be written as 

\begin{equation}
     \tau(\bar{\nu}) = \tau_{\mathrm{max}} \exp\bigg(-\frac{4\ln(2)(\bar{\nu}-\bar{\nu}_0)^2}{\Delta_{\bar{\nu}}^2}\bigg)
\end{equation}

and

\begin{equation}
    \tau_{\mathrm{max}} = \rm N \times A_{\mathrm{mode}} \times \frac{2\sqrt{\ln(2)}}{\Delta_{\bar{\nu}} \sqrt{\pi}}
,\end{equation}where $\tau_{\mathrm{max}}$ is the absorption maximum in optical depth, $\bar{\nu}_0$ the vibration frequency, $\Delta_{\bar{\nu}}$ the FWHM (cm$^{-1}$), and A$\rm_{mode}$ the integrated band strength (cm/molecule).
The absorption spectrum can then be obtained by using $\exp(-\tau(\bar{\nu})$).\\
The inverse operation to compute the column density is then

\begin{equation}
N=\frac{1}{A_{\mathrm{mode}}}\int_{band}{\tau_{\nu} \ d\nu}, 
\label{eqn}
\end{equation}
where $N$ is the species column density (in cm$^{-2}$), and $\tau_{\nu}$ is the optical depth of the absorption band.\\

For large absorption features that need to be fitted by a sum of Gaussian profiles, the initial band strength measured in the lab is divided for each component with a least-squares method. As of now, we do not consider the effect of small particle shapes, such as CDEs \citep[see e.g.][]{Imai_2009} on the absorption profiles.

\subsection{Database and fitting}

The present version of SynthIceSpec is based on the fitting of pure ice spectra measured in the laboratory. The different spectral databases used are the Optical Constants Database (OCdb)\footnote{\url{https://ocdb.smce.nasa.gov/}}, the Leiden Ice Database \citep{Rocha_2022}\footnote{\url{https://icedb.strw.leidenuniv.nl/}}, the Cosmic Ice Laboratory,\footnote{\url{https://science.gsfc.nasa.gov/691/cosmicice/}} and  the Solid Spectroscopy Hosting Architecture of Databases and Expertise (SSHADE)\footnote{\url{https://www.sshade.eu/}}.

\begin{figure*}
    \centering
    \includegraphics[width=0.85\linewidth]{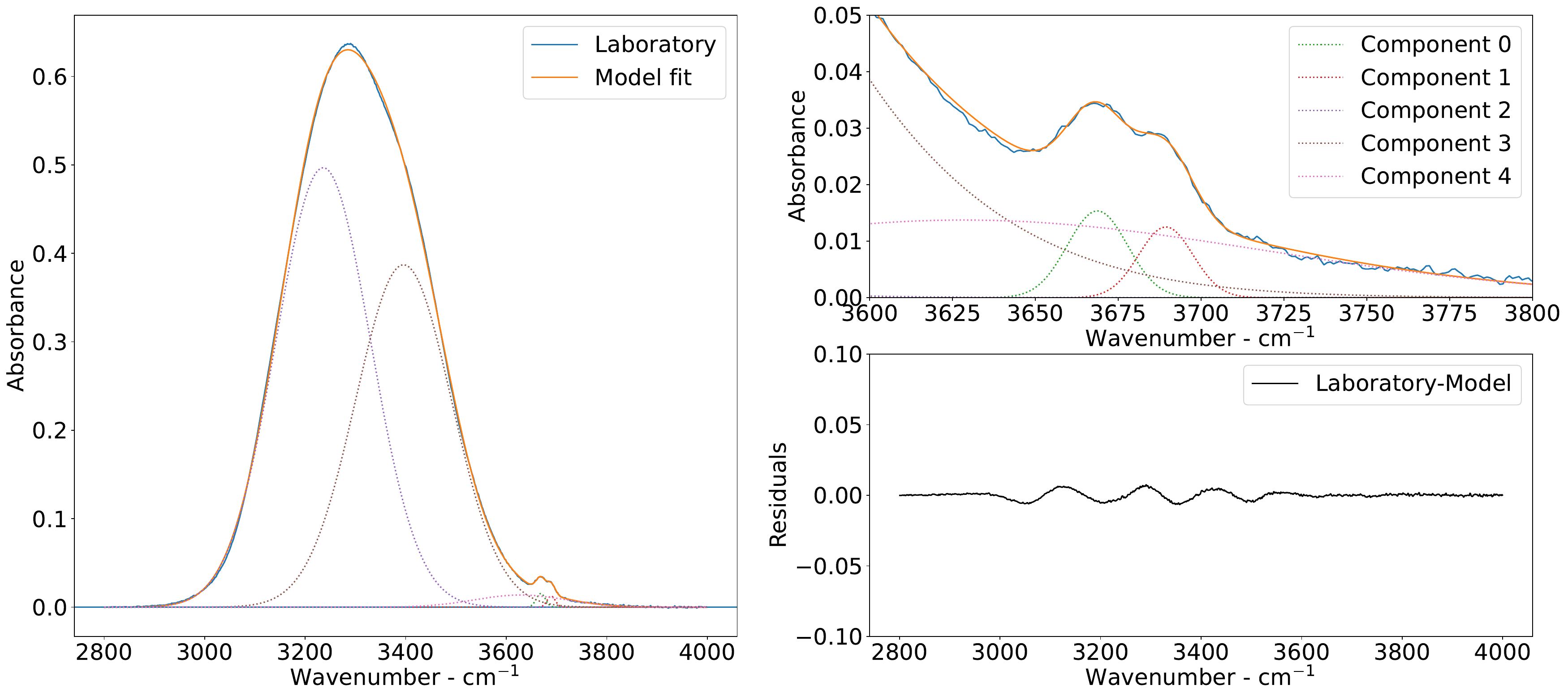}
    \caption{Our Gaussian fitting (in orange) of the 3$\mu$m absorption feature of pure H$_2$O from \citet{Oberg_2007} (in blue) as a function of the wave numbers. The top right panel focuses on the dangling OH of water. The bottom right panel shows the residuals between our fit and the laboratory spectrum. }
    \label{fig:pure_water_fitting}
\end{figure*}

For each vibrational mode, the fitting is done directly from a laboratory spectrum of the considered molecule using a least-squares fitting. The number of Gaussian components is set by the minimum number of Gaussians needed until the residuals left after subtracting the fitting to the laboratory spectrum is lower than 3 RMS. All the parameters (position, bandwidth, and band strength) derived from these fits can be found in Appendix.~\ref{appendix-data}, alongside the references associated with each species. Note that these parameters may not be suitable for fitting other laboratory spectra, as the position of the bands may vary slightly, depending on the baseline subtraction and/or the spectroscopic technique used in the experiment.
We started implementing the latter components \citep[see e.g.][]{Taillard_2025}, but the mixed ice database will be released at a later date. 
In Fig.~\ref{fig:pure_water_fitting} we show an example of our Gaussian fitting for the 3 $\mu$m water feature on the  pure spectra at 10 K measured in the laboratory by \citet{Oberg_2007}. We fitted a series of Gaussian profiles to better reproduce the shape of the feature, including the dangling O-H modes. The laboratory spectrum is shown in blue, where we sum the multiple components, resulting in the orange line. In the bottom right panel, we show the residuals between our fit and the laboratory spectrum. 

Any user can add their own Gaussian parameters of any species directly in the database file.
All these fitting parameters and the future ones that will be implemented are all listed in a public database named IceSpecData\footnote{\url{https://astrochem-tools.org/}}, which is accessible online or directly from the code's files.  

\subsection{Instruments and noise}

We have implemented three different JWST instruments: NIRCam, NIRSpec, and MIRI. Each instrument has its own resolving power and wavelength range that is directly extracted from the Space Telescope Science Institute  documentation (STScI)\footnote{\url{https://jwst-docs.stsci.edu}}. 
The filters for NIRSpec were implemented and in the MIRI case, we included the strongly wavelength-dependent resolving power in four different modules as follows: 4.88-7.52 $\mu$m with R$\sim$3500; 7.52-11.75 $\mu$m with R$\sim$2900; 11.75-18.08 $\mu$m with R$\sim$2200; 18.08-28.34 $\mu$m with R$\sim$1700. 
NIRCam only has the GRISM option for now.
Each filter can be selected individually, or a ``full'' template can be used. 
The filter's throughputs were not taken into account at this point and were considered constant in the wavelength range for each filter of each instrument. 
When unavailable, the resolving power and wavelength grids are implemented with an average value on the entire wavelength coverage and will be updated as the technical data are released. An arbitrary grid can also be used and defined by the user (examples are provided within the code).
It is also possible to add intrinsic instrumental noise in the generated spectra, either manually or from a template.

\subsection{Dust grain and extinction}

The interstellar grains are responsible for the scattering of the light in the ISM leading to an extinction effect in particular in the near-IR. At this time, we were not able to fully reproduce the diverse size and shape effects at play; hence, both the ice and refractory parts of grains were parametrised considering a flat, layered geometry. We detail the assumptions we used in the following sections.

\subsubsection{Silicates implementation}\label{sec:silicates}

Silicates (or astrosilicates) are extremely important in the ice observations since they contribute to strong absorption bands (e.g. at 9.7 $\mu$m). Optical properties of silicate grains were defined in \citet{Draine_1984} based on the MRN models from \citet{Mathis_1977} and have been widely studied since then (\citet{Demyk_2022,Hensley_2023,Ysard_2024}). 
Our approach does not currently consider the possible shape of ice-coated dust grains, as they are observed in the ISM, where the exact geometry distribution and composition are not known \citep{Fabian_2001,Min_2003,Min_2007,Min_2008,Koike_2010}. 
Multiple models have approached the effect of geometry on the scattering of the light; for example, the continuous distributions of randomly oriented ellipsoids (CDEs) \citep{Bohren_1983,Zubko_1996,Draine_2021b,Draine_2021a} or the discrete dipole approximation (DDA) \citep{Draine_1994,Draine_2008,Draine_2013,Dartois_2022}. 
In SynthIceSpec, we only considered a flat geometry, as the laboratory spectra we used were produced by applying layers of ice to a flat surface. 
The addition of the geometry effect to the calculated extinction will be considered in the future.

We have so far included silicate signatures that can be tuned depending on the hydrogen column density and dust-to-gas mass ratio input by the user. 
The silicate template is computed from the composition of the so-called astronomical silicates' optical constants from \citet{Li_2001}, giving the mass absorption coefficient (cm$^2$/g) of a thin silicate film.
This coefficient is converted into optical depth assuming a refractory dust grain column density, N$\rm_{grains}$. The latter is computed assuming that N$\rm_{grains}$ is equal to the ratio of the hydrogen column density (input by the user) over the gas-to-dust mass ratio expected in the target (with the standard value being 100).
The optical depth is then computed as

\begin{equation}
    \tau = (\kappa_{abs}+\kappa_{sca}) \times N_{grains}  \times m_{grain} 
.\end{equation}
Here, $\kappa_{abs}$ and $\kappa_{sca}$ (cm$^2.g^{-1}$) are, respectively, the absorption and diffusion coefficients. N$_{grains}$ is the column density of the grains (cm$^{-2}$), and m$_{grain}$ is the grain mass (g).
For a pure extinction, the observation can be deduced by $\exp{(-\tau)}$, meaning that the resulting absorption spectra from the ices and silicates can be obtained using a first-order approximation of $\exp{(\tau + \tau_{ice})}$. 

The silicates spectrum reproduces the features found in the translucent interstellar medium \citep[e.g.][]{Dorschner_1995,Kemper_2004}. As silicates are not prone to evolving and changing their composition in cold environments \citep{Fabian_2001}, we used this spectrum as an input for all the simulations of the type of source presented hereafter. In doing this, we intrinsically assume that class 0 and class I are not warm enough to modify the silicates underlying structure. Silicates do, however, present a short wavelength wing on the 9.7 $\mu$m feature that is not present in the diffuse medium, caused by different dust properties \citep{Gao_2010,van-Breemen_2011,Boogert_2011}, but is present in our silicate model.

\subsubsection{Extinction law}\label{sec:extinction_law}

Background star observations are affected by the extinction caused by the absorption and scattering of photons by dust grains \citep{Draine_2003}. 
For now, we do not reproduce the extinction taking into account the shape and size of the grains in SynthIceSpec. 
We thus used an empirical extinction model. 
In an extensive study of ices in cold cores, \citet{Boogert_2011} derived a formalism to determine the extinction curve of their targets after removing both the ice and silicate features in their observations, providing a continuum-only extinction law. By deducing the stellar type of the background source using their models, the authors described the extinction as

\begin{equation}\label{eq_extinction_curve}
    A_\lambda = -2.5\; log \; ( F_\lambda(obs)/F_\lambda(model))
,\end{equation}where $F_\lambda(obs)$ and $F_\lambda(model)$ are, respectively, the observed flux and the modelled flux of the star. Furthermore, the authors derived a polynomial fit, P, from a feature-free region, depending on the extinction in the K band, A$_K$. This polynomial extinction curve takes the form of

\begin{equation}\label{polynomial_ext_cuve}
    log \; ( A_\lambda/A_K) = a_0 + a_1\;log(\lambda) + a_2\;[log(\lambda)]^2 + ...
,\end{equation}
with $\lambda$ in $\mu$m and A$_\lambda$ in magnitude. The polynomial coefficients are available in Table 4 of \citet{Boogert_2011}. The curve is rather flat in the mid-IR, corresponding to R$_V$ $\sim$ 5 \citep{Gordon_2023}. 
To add this feature to SynthIceSpec, we applied the opposite formula. Here, the unknown parameter is the $F_\lambda(obs)$, while the stellar type and visual extinction, A$_V,$ are given by the user. We computed the A$_K$ from the conversion A$_K$ = 0.112 $\times$ A$_V$ \citep{Johnson-Morgan_1953}. We then applied the polynomial, P, to the A$_K$ as so:

\begin{equation}
    A_{\lambda,ext} = 0.112 \times A_V \times 10^{P}
,\end{equation}leading to the following observed stellar flux:

\begin{equation}\label{eq-extinction}
    F_\lambda(obs) =  F_\lambda(mod) \times e^{\frac{-A_{\lambda,ext}}{2.5}}
,\end{equation}which was applied directly to the ice-absorption spectrum. 

\subsection{Stellar spectra continuum and modified black bodies}

By default, if no stellar continuum is adopted, the flux appears as constant on all wavelengths. If a stellar spectrum is used, the resulting ice spectrum will be the combination of the absorption spectrum and the stellar one. We offer multiple options of synthetic continuum.

\subsubsection{Stellar spectra continuum}

SynthIceSpec can take a stellar spectrum as input to reproduce the photospheric absorption bands. The photosheric absorption bands and continuum are linked to a few physical parameters such as the star surface temperature, the surface gravity, or the metallicity. 
To meet our needs, we used the PHOENIX models and its extensive database \citep{Allard_1995,Hauschildt_1999,Gustafsson_2008,Husser_2013} to produce the photospheric spectra, adjusted to the instrument's resolution and wavelengths. The PySynphot\footnote{\url{https://pysynphot.readthedocs.io/}} Python library allows us to access the spectra catalogue. The user can then choose to generate their stellar continuum and the emerging spectrum with ice absorptions through three different parameters to define the stellar type: effective temperature of the star (T$\rm_{eff}$), surface gravity (log(g)), and metallicity ([M]/[H]). It is also necessary to input the distance to the star (the default is 1 kpc) and its radius (the default is 0.5 R$_{\odot}$).

\subsubsection{Modified black bodies}

\begin{figure}
    \centering
    \includegraphics[width=0.7\linewidth]{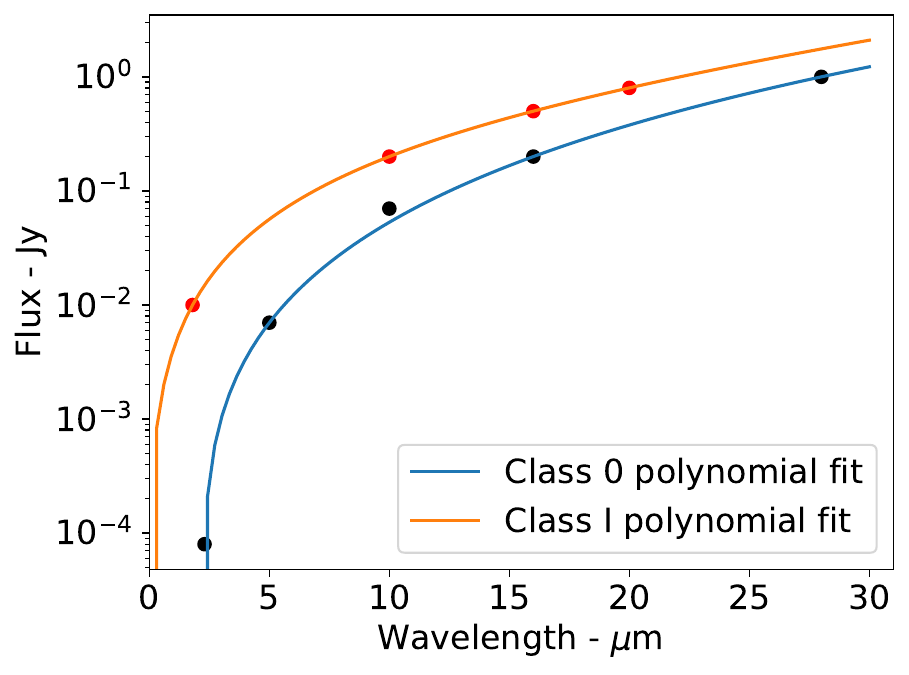}
    \caption{Polynomial fits to the continuum emission from extracted data points from \cite{Yang_2022} and \cite{Tychoniec_2024} for a class 0 and class I protostar respectively.}
    \label{fig:polynomial}
\end{figure}

It is possible to observe young stellar objects (YSOs) with the JWST, and a few spectroscopic observations have already been published. In \citet{Yang_2022}, when studying a class 0 object (IRAS15398–3359) the authors derived a fourth-order polynomial to estimate the large-scale continuum and divide the flux density by the resulting continuum to derive the optical depth of the ice features. Here, we used their fit and picked five points in order to recreate our own polynomial, mimicking the modified black body the authors derived.
Similarly, we used the spectrum published in \citet{Tychoniec_2024} of a class I protostar, TMC-1 W, with only four points this time.
These two polynomial fits can constitute synthetic continua. We present both polynomial fits in Fig.~\ref{fig:polynomial}, with class 0 in blue and class I in orange. 
These are fiducial continua and are included for the user to anticipate a possible shape of the continuum. All YSOs present their own continua that should be studied case by case as they depend on multiple parameters intrinsic to the observation itself \citep{Myers_1993,Nakajima_1997,Garufi_2018}; these include geometry of the object, envelope, inclination, bolometric temperature, and so on.
The templates can be used to carry out exploratory projects that need to assume a `standard' class 0 or class I Spectral Energy Distribution (SED) to study, for example, the detectability of a given species in different environments \citep[see][]{Taillard_2025}. The user can also define or use their own preferred SED.

In a future update, we will make it possible to input any profile from the intensive database of YSO SEDs originally published in \citet{Robitaille_2006} and its new versions \citep{Robitaille_2007,Robitaille_2017}. In Appendix.~\ref{appendix-decomposition} we decompose the different elements present in SynthIceSpec added to an ice-absorption spectrum.

\subsection{Gas-grain model outputs}

One of the important aspects of the tool is its use in the post-processing of gas-grain models to predict the observed ice spectra. 
It only needs to retrieve the different species and abundances predicted by any gas-grain model and use them directly in the code. We present examples of how to use SynthIceSpec with models in Section.~\ref{use_case}.
In Appendix.~\ref{appendixB}, we show an example of how SynthIceSpec can trace the ice evolution through time, with chemical model outputs at four different time steps. The resulting spectra for each are plotted in different colours (from blue to red as time increases), allowing us to follow the evolution of the ice composition and see the switch from H$_2$O-dominated ice to CO$_2$-dominated ice.

\subsection{Limitations of the tool}\label{sec:limitations}

SynthIceSpec was built on a simple approximation. The parameters listed in the public database were derived from pure laboratory spectra. This means that mixed ices as a whole are not yet implemented. This simplification is the biggest weakness of the tool, because it may affect the position, shape, and strength of a band. In addition, the code assumes a uniform average ice composition along the LOS, instead of a more realistic scenario including different layers and dust populations along the cloud.

A certain number of species are also missing from the IceSpecData database, mostly from COMs found in the fingerprint regions (1500 to 500 cm$^{-1}$). As new IR ice spectra become available, we will post-process them and add their parameters to IceSpecData.

\begin{figure}[!h]
    \centering
    \includegraphics[width=0.8\linewidth]{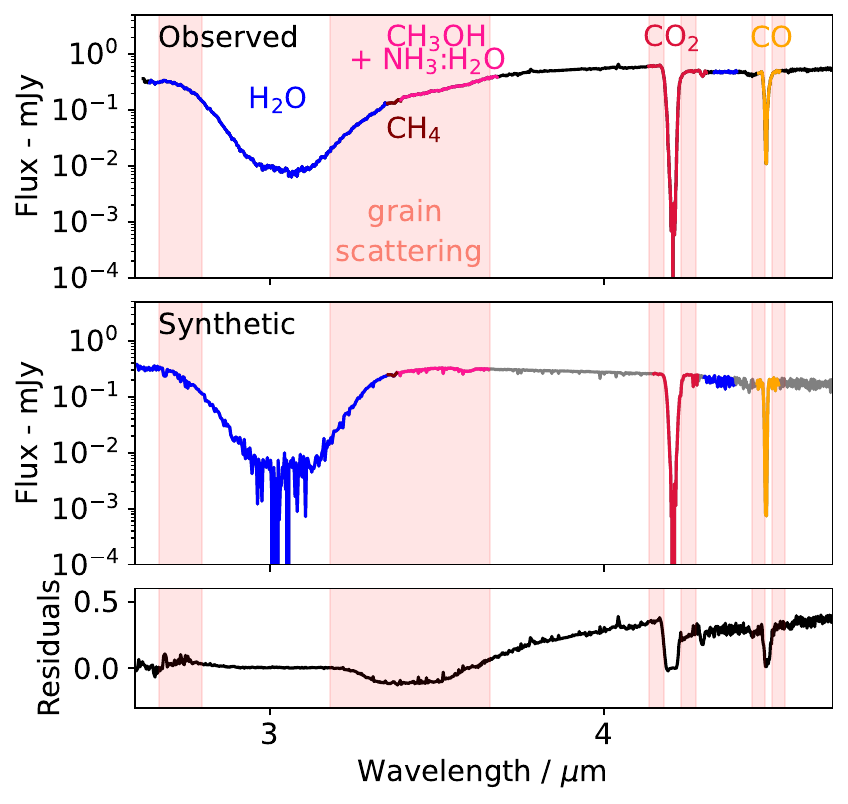}
    \caption{Comparison between JWST observation of NIR38 background star (top plot) in Chameleon I \citep{Mcclure_2023} with NIRSpec FS/395H and the synthetic spectrum (middle plot) obtained with the same column densities derived by the observers, with NIRSpec FS/395H parameters. The stellar spectrum applied is from a K7V star, as it was determined in \citet{Dartois_2024} as a good match to the observed photospheric lines. The absolute residuals (observations minus the synthetic spectrum) are given in the bottom plot. The pink rectangles indicate where the grain growth effect is notable in the observation.}
    \label{fig:comp_mcclure}
\end{figure}

The Gaussian parameters used were extracted from spectra at a low temperature, meaning that certain physical processes affecting the bands are not present (e.g. crystallisation). Data from a thermally processed molecule presenting behaviour such as crystallisation or segregation are not included. 
Grain size, aggregation, and grain-shape effects have been widely studied and modelled
\citep{Weingartner_2001,Zubko_2004,Ormel_2009,Paruta_2016,Jones_2017_THEMIS,Dartois_2024}, and we will try to parametrise some of these effects in future developments of the code. In Figure.~\ref{fig:comp_mcclure}, the top panel shows the observations made by \cite{Mcclure_2023} towards the background star NIR38 and in the middle panel, the synthetic spectrum obtained with the same column densities and stellar type derived from the IceAge study. The last panel shows the absolute residuals of the subtraction between the observed and synthetic spectra. As mentioned, the grain growth, illustrated by the pink rectangles in the two upper panels, is not yet implemented, and the difference is noticeable. Another disagreement arises for the CO and CO$_2$ bands, which are probably affected by mixing and grain growth (in the case of CO$_2$ in particular). 
Grain-growth models' outputs or parametrisation could be easily implemented in SynthIceSpec and lead to a better reproduction of the observations. 

In this state, SynthIceSpec can still be used as a proxy to identify and compare the position of certain bands. It can also be used to programme observations from model results and help identify unknown species on actual data. This was illustrated in \citet{Taillard_2025}, where a detailed study of the detectability of S-bearing species is presented. It is highly flexible and any new laboratory spectrum of any species could be easily added to the database, which will be updated yearly as spectroscopic measurements become available. We are also planning to make a similar database for mixed ices available. New stellar templates and more accurate class 0 and I SEDs will be included in the code database as delivered from new JWST observations.
Parallel to updating IceSpecData, we will continue to develop new features to facilitate analyses and predictions.

\subsection{Summary}

Compared to tools such as the JWST exposure-time calculator (ETC), we provide a more flexible option that can provide extinction at the same time as its own ice composition. In the context of ice and ISM observations, the ETC alone can provide a basic analytic spectrum (flat and power-law continuum, black-body spectrum) or a stellar spectrum (from models such as Phoenix), to which an extinction law from different formalisms can be applied (e.g. Weingartner \& Draine 2001). The biggest difference is the versatility of SynthIceSpec, where the user can  easily change the ice and dust compositions. The user can also include their own continuum template and extinction law.

SynthIceSpec can be summarised as this routine computing the synthetic flux I$\rm_{synth}$ as I$\rm_{synth}$ = I$_0$ $\times$ $exp(-\tau),$ where I$_0$ is the flux of the background source and $\tau$ the optical depth of the source. Next, the following workflow is applied following the user's choice:

\begin{itemize}
    \item The user chooses the instruments (NIRSpec, NIRCam...) and filter (the full wavelength range for each instrument is also available).
    \item The user inputs the background source: flux (standard is 1 mJy), ice composition, and visual extinction (should an extinction be desired).
    \item The optical depth $\tau$ is then computed based on the sum $\tau$ = $\tau_{ext}$ + $\tau_{silicates}$ + $\tau_{ice}$ and corresponding to the following: $\tau_{ext}$, where extinction is computed from the initial A$_V$ using a \citet{Boogert_2011} theoretical continuum-only extinction curve. We then calculated $\tau_{silicates}$, to which silicate features can be added (as a function of N(H$_2$)) from a  flat or spherical geometry model. Finally, we computed $\tau_{ice}$ (i.e. ice absorption) from the initial composition.
    \item The synthetic spectrum is generated as I$\rm_{synth}$ = I$_0$ $\times$ $exp(-\tau)$ at the resolution and wavelength range of the chosen JWST instruments. I$_0$ is the continuum that can be tuned as well, for which different options are available: flat, stellar (from Phoenix models), and black body (simple, modified, and templates of YSOs are available).
\end{itemize}

Compared to tools similar to the ETC of the JWST, ours is more flexible and can provide extinction at the same time as ice composition. In the context of ice and ISM observations, the ETC alone can provide a basic analytic spectrum (flat and power-law continuum, black-body spectrum) or a stellar spectrum (from models such as Phoenix), to which an extinction law from different formalisms can be applied \citep[e.g.][]{Weingartner_2001}. The biggest difference is thus the versatility of SynthIceSpec, which allows the user to  easily change the ice and dust compositions. The user can also include their own continuum template and extinction law.

\section{Use cases}\label{use_case}

In this section, we present two use cases of SynthIceSpec. We first considered the post-processing of chemical modelling from \citet{Taillard_2023}, with two `extreme' grain-temperature cases, and we ran new models with intermediate physical parameters. We then compared the synthetic spectra derived from those models to the IRTF and \textit{Spitzer} observations published in \citet{Boogert_2011}. 
We probed the effect of dust temperature in Nautilus gas-grain model and study its impact on CO$_2$ formation. Here, we present how SynthIceSpec can be used to determine the detectability thresholds of new species, using CH$_3$CN as an example. We also warn the reader of the impact of photosphere absorption lines that can hinder the detection of molecules.

\subsection{Post-processing of chemical modelling}\label{CO-study}

SynthIceSpec can be applied to any kind of chemical modelling outputs, such as post-processing, to predict ice observations. The user, however, needs to convert the modelled abundances into column densities before inputting them in the code. A way to do it at first order is to use the visual extinction input in the model parameters and convert into N$\rm_{H_2}$ with the factor N$\rm_H$ = 1.8 $\times$ 10$^{21}$ $\times$ A$_V$ \citep{predehl_x-raying_1995,ryter_interstellar_1996,olofsson_extinction_2010}, before dividing by two to obtain the N$\rm_{H_2}$ column density. As such, the modelled abundances (with respect to H) can be converted into column densities (with respect to H$_2$) and input in SynthIceSpec. In this section, we focus on the impact of grain temperature on the formation of ices and the resulting spectra. 

\subsubsection{Static-chemical-model description}

\begin{table*}[]
    \centering
    \caption{Ice column densities.}\label{tab:col_ice}
    \begin{tabular}{ccccccc}
          \hline
          \hline
          Name               & N(H$_2$O)                     &  N(CO$_2$)                     & N(CH$_3$OH) \\
                          &($\times$ 10$^{18}$ cm$^{-2}$) & ($\times$ 10$^{17}$ cm$^{-2}$) & ($\times$ 10$^{17}$ cm$^{-2}$) & \\
          \hline
          \hline
          2MASS J18170957-0814136       &  2.85 (0.31)                  &  12.29 (1.29)                  &  2.69 (0.58)     \\
          \hline
          Model - T$\rm_{dust}$ = 6.6 K                  &  14.7                           &  0.1                             &  4.6   & \\
          Model - T$\rm_{dust}$ = 10.3 K                &  6.7                            &  16.9                           &  0.7   & \\
          Model - T$\rm_{dust}$ = 13.2 K                &  6.1                             &  139.6                            &  4.3    &\\
          \hline
 
    \end{tabular}
    \tablefoot{Column densities observed towards the background star 2MASS J18170957-0814136 as reported in \citet{Boogert_2011}, with the uncertainties in parentheses, and the predictions of the three models described in Section.~\ref{CO-study}.}
\end{table*}

\begin{figure}[]
    \centering
    \includegraphics[width=0.8\linewidth]{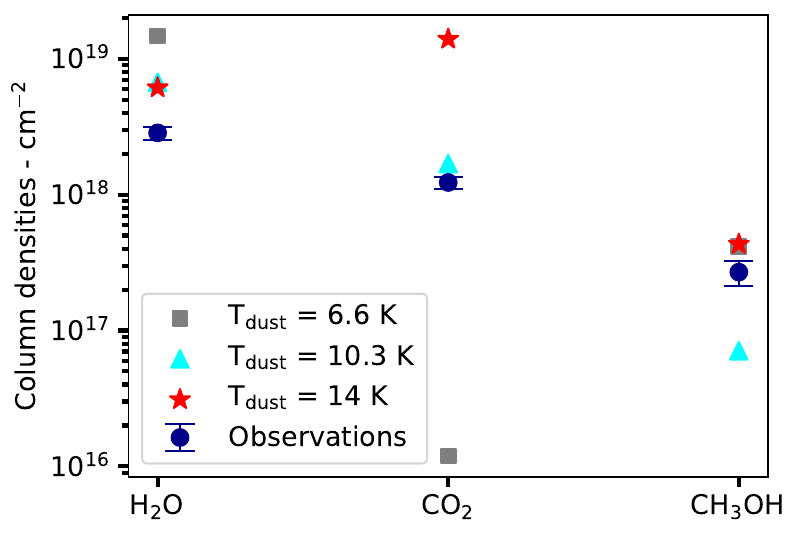}
    \caption{Observed column densities for H$_2$O, CO$_2$ and CH$_3$OH and their uncertainties as derived in \citet{Boogert_2011} towards the background star 2MASS J18170957-0814136, shown in dark blue.
    We also plot the column densities predicted by three different models aiming to reproduce the observed CO$_2$ column density. Except the dust temperature, they all share the same physical parameters retrieved from \textit{Herschel} data of the cold core L429-C (see text). Grey squares correspond to a model with the dust temperature computed with the parametrisation from \citep{hocuk_parameterizing_2017} with T$_{dust}$ = 6.6 K; cyan triangles correspond to a model with a fixed temperature of T$_{dust}$ = 10.3 K, and red stars correspond to a dust temperature retrieved from \textit{Herschel} data, with T$_{dust}$ = 14 K. }
    \label{fig:best_models}
\end{figure}

In \citet{Taillard_2023}, the authors studied the chemical composition of the cold core L429-C, as it was the first core where methanol ices were observed with \textit{Spitzer} \citep{Boogert_2011}. The presence of high column densities of solid methanol reflects an advanced chemistry, which was confirmed by gas-phase observations and the CO freeze-out occurring in the source.
Following \citet{Taillard_2023}, the authors used Nautilus \citep{Ruaud_2016} to compare both in gas and solid phases) to the model outputs to determine, within a grid of physical parameters, which ones best reproduced the observations. In this instance, the dust-temperature parametrisation, derived from visual extinction obtained from \textit{Herschel} data, presented in \citet{hocuk_parameterizing_2017}, was used to allow a better reproduction of the solid methanol and gas phase rather than using a higher dust temperature. 

In previous studies, it was shown that the gas temperature impacts the chemistry of both phases \citep{Clement_2023,Taillard_2025b}. In the present paper, we focused on how the dust temperature deeply affects the formation of ice species such as CO$_2$. To do this, we used the solid-phase observations of Keck combined to \textit{Spitzer} towards the cold core L429-C \citep{Boogert_2011} and investigate the consequences of changing the dust temperature in model predictions. We were particularly interested in the background star 2MASS J18170957-0814136 (hereafter J18170957). It is the only object towards which the long-low (14-21.3 $\mu$m) module was used, allowing the observation of the CO$_2$ bending mode at 15.0 $\mu$m.
The physical parameters are shared in all the models and were derived following the method used in \citet{Taillard_2023} at the star position. The parameters are
A$\rm_V$ = 49.7 mag; n$\rm_H$ = 4.5 $\times$ 10$^5$ cm$^{-3}$; and cosmic-ray ionisation rate: $\chi_{CR}$ = 3.00 $\times$ 10$^{-17}$ s$^{-1}$. A single grain radius was considered here (r$\rm_{grain}$ = 0.1 $\mu$m).
The initial abundances are the ones presented in \citet{Wakelam_2021}, and we used the new upgraded chemical network for Nautilus \citep{Wakelam2024}. During our results analysis, we noticed that increasing the H volume density number only slightly impacts the CO$_2$ abundance; this is due to the fact that in all models the core is already experiencing CO depletion, and this only slightly shifts the time at which the models best reproduce the observations. As the density was high, we assumed that the dust and gas temperatures were coupled. 

First, we ran two static models: the `cold model' using the \citet{hocuk_parameterizing_2017} parametrisation and the `warm model' using the temperature derived from \textit{Herschel} data. 
The cold model has a dust temperature of 6.3 K, whereas the warm model's is 13.2 K, representing two extreme cases. The best model (i.e. the one where the observed abundances are the closest to the modelled abundances with respect to time) was determined using the so-called distance of disagreement defined in \citet{Wakelam_2006} as the fitting parameter. The best times are t$\rm_{cold}$ = 3.5 $\times 10^{5}$ yr and t$\rm_{warm}$ = 2.7 $\times 10^{5}$ yr.
After reviewing the results (see below), we ran a set of models with the dust temperature varying between 9 and 13 K. For each model, we compared the different abundances to the CO$_2$ one, which we show in Appendix.~\ref{dust-models}. We iteratively modified the dust temperature until we reached a good agreement with the CO$_2$ column densities.
We then took into consideration this third model as `intermediate', with a dust temperature of 10.3 K. The `best time' for this model is t$\rm_{intermediate}$ = 2.7 $\times 10^{5}$ yr.

For all three models (cold, intermediate, and warm), we extracted the abundances, converted them into column densities, and used them as inputs for SynthIceSpec; these are reported in Table.~\ref{tab:col_ice} along with the observed column densities. The values were then used to compare the different models and their associated chemistry.
The results are also reported in Fig.~\ref{fig:best_models}. For the ice species H$_2$O, CO$_2,$ and CH$_3$OH, which were observed towards the background star J18170957, we plot the different column densities as derived in \citet{Boogert_2011} (in dark blue) and their predicted values for each model (grey square for the cold model; cyan triangle for the intermediate model; red star for the warm model). Unfortunately, the CO stretching-mode feature at 4.7 $\mu$m was not within the observational range and was not studied in this work as we did not obtain proper measurements of its column density, despite it being the most important link between CO and CO$_2$.

\subsubsection{Dust-temperature impact on CO$_2$ chemistry}

In all models, H$_2$O is overproduced by a factor between three and seven, with the warm model being the closest to the observations and the cold model the furthest. CO$_2$ is highly dependent on the model grain temperature; this is because it allows the diffusion of CO radicals at a certain threshold. The best model is the intermediate one, while the warm model overproduces the column density by a factor of ten, and the cold model is two orders of magnitude below the observation. Finally, CH$_3$OH is well reproduced in the cold and warm models but under-produced by approximately a factor of four by the intermediate model; this is because most of the CO is converted in CO$_2$. From Table~\ref{tab:col_ice} and Figure.~\ref{fig:best_models}, we see that there are no models that can simultaneously reproduce all three species, leaving an erroneous elemental budget for the rest of the simulations.

We can illustrate the impact of dust temperature in the models by looking at by far one of the most impacted molecules, CO$_2$. In the cold model, the formation of the molecule is indeed inefficient as both the CO and O diffusion energy barriers are too high to effectively form CO$_2$. In this case, when the radicals are locked, we see an increase of H$_2$O column density because H atoms can easily move across the grain surface. When the grain temperature is high enough, the radicals can move freely and react to drastically increase CO$_2$ abundance. This was illustrated in recent modelling studies \citep{Clement_2023,Jimenez-serra_2025}, where implementing a higher temperature ($>$ 12 K) led to strong changes in the ice composition and a more CO-dominated chemistry.

Since dust temperature is crucial in solid-phase chemistry, it must be strongly constrained, and multi-wavelength studies are required to observe dust of different size distributions (in emission and absorption).
Indeed, as already shown in various studies, using \textit{Herschel} data to probe temperature can incur significant limitations as it only one grain population is covered \citep{Bernard_2010,Hirashita_2011}. Using the analytical expression derived in \citet{hocuk_parameterizing_2017} can also be expected to have fairly strong uncertainties linked to the A$_V$ determination and the cold model is the one showing the biggest discrepancy compared to the observations. It is then necessary in models to consider grain temperature dependency on the grain radius. As such, adding a grain size distribution (with each size having its own temperature), a 1D radial gradient of the temperature as we go deeper in the cloud, or a better constraint on the time evolution as grains aggregate and become colder with increasing density could lead to a better agreement of the abundances of main ice species in models with those observed. We had to rely on observations to better constrain the dust temperature before introducing new paradigms in models.
Deriving the dust temperature using gas-phase observations remains tricky because of opacity effects, but certain key species can be used to obtain the kinetic temperature with high precision \citep[e.g. with NH$_3$ in][]{Lin_2023}.
From this perspective, observations of solid CO$_2$ can provide additional constraints on the dust temperature via the column of CO$_2$ formed and the information on grain size and shape distribution available in the scattering wings. Of course, consideration on the shape of the ice-coated dust grains are also needed to achieve a higher degree of precision on the dust temperature as the diffusion of chemical species on the surface is also highly dependent on this parameter.

Grain sizes can be probed by scattering effects, which become significant when the grain size is $\sim$ $\lambda$/2$\pi$. It strongly affects the vibration modes at short wavelengths. It was recently studied in \citet{Dartois_2024} at near-IR, where the authors modelled the grain growth observed in two background stars towards Chamaeleon I using JWST/NIRSpec. The CO$_2$ stretching-mode absorption feature at 4.2 $\mu$m in particular can be used to derive information on the size distribution of grains probed along the LOS, making it a great target of which we can add constraints to both the dust temperature and grain size. 
Solid CO$_2$ observations can therefore be of paramount importance to strongly constrain the physical property of dust and its impact on ice chemistry.
Observations could be made with the JWST to carry out in-depth studies within the same clouds with the aim of understanding how dust properties impact the overall ice composition. As was demonstrated in a recent study \citep{Smith_2025}, the JWST is capable of mapping the most abundant ice constituents, including CO$_2$, in deeply extinguished regions.

In the bigger picture, CO abundance on the grain is thus key to driving a CO-rich chemistry. In models, the CO ice abundance itself is a tight constraint on the dust temperature and is the precursor of CO$_2$ through O addition. The current description of diffusion in Nautilus is entirely linked to the dust temperature, and more constraints must be added in order to more realistically reproduce interstellar chemistry.
Chemical models must therefore include constraints within their chemical networks, formalisms, and mechanisms where numerous other parameters must be taken into account to truly probe the impact of temperature on CO$_2$ formation. These mechanisms and parameters include, for example, depletion time \citep{Taillard_2025b}, sticking coefficients \citep{Stadler_2025}, grain-size distribution \citep{Iqbal_2018}, diffusion mechanisms \citep{Minissale_2013,Clement_2023}, and binding energies \citep{Noble_2012,Molpeceres_2023,Navarro-Almaida_2025}.

\subsubsection{Comparing observed and synthetic spectra}

\begin{figure*}[!h]
    \centering
    \includegraphics[width=0.6\linewidth]{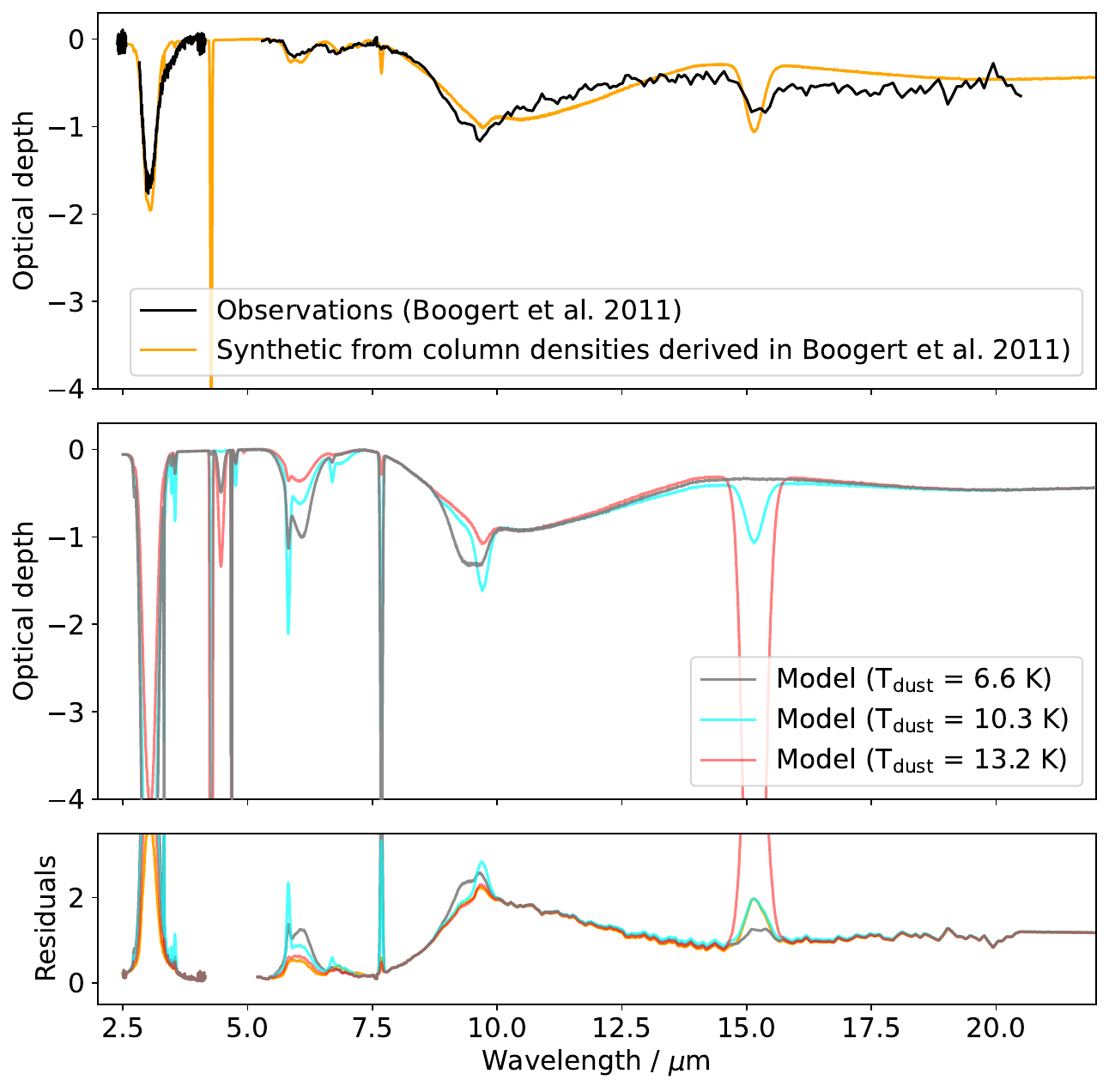}
    \caption{
    Top: Optical depth derived from observation towards 2MASS J18170957-0814136 from \citet{Boogert_2011} (black) and the synthetic optical-depth spectrum using the column densities derived from the latter (orange).
    Middle: Synthetic optical depth from Nautilus predictions; the `cold' model where the dust temperature (T$\rm_{dust}$) was computed with the parametrisation from \citet{hocuk_parameterizing_2017} (T$\rm_{dust}$ = 6.6 K) is shown in grey; the `intermediate' model (T$\rm_{dust}$ = 10.3 K) is shown in cyan; the 'warm' model where T$\rm_{dust}$ was obtained from \textit{Herschel} data (T$\rm_{dust}$ = 13.2 K) is shown in red.
    Bottom: Absolute residuals from subtraction between observations and synthetic spectra presented above.
    }
    \label{fig:Figure_L429C}
\end{figure*}

In Figure.~\ref{fig:Figure_L429C}, we show the observed spectrum of J18170957 in black in the upper panel in optical depth as a function of wavelength (in microns). In the middle panel, the synthetic spectra from the three different models (the cold model is in grey; the intermediate model is in cyan; the warm model is in red) are plotted for comparison. 
The intermediate model, as it is closest to the observed CO$_2$ column density, has the lowest residuals between the observed and synthetic spectra around 15 $\mu$m.
The warm model, with the H$_2$O column density being the lowest of the three, reproduces the water feature around 6 $\mu$m, the methanol feature next to it at 6.8 $\mu$m, and the other small features (e.g. CH$_4$ feature at $\sim$ 7.6 $\mu$m and H$_2$CO feature at $\sim$ 6.7 $\mu$m) well.
However, the latter, small features are very strong in the cold and intermediate models' synthetic spectra. With CO$_2$ in the warm model, these species are overproduced (2-50 times more) by the model. These three different spectra show how a simple change of a few kelvin at the surface in chemical models could strongly impact the production of molecules and, thus, their absorption features.

We compared the observed optical-depth spectrum to the ones from the chemical model outputs to better understand how the two differ; this is shown in the lower panel of Figure.~\ref{fig:Figure_L429C}.
Starting with the water band around 3 $\mu$m, we can see that, as mentioned before, the three models are overproducing it. This is a known problem in Nautilus; depending on the dust temperature used, oxygen is either be massively locked in water or CO$_2$ \citep{Clement_2023}. The modelled abundances are therefore much higher than what \citet{Boogert_2011} observed, resulting in a deeper absorption and higher residuals. However, as for the orange spectrum using the column densities derived from observations, it is very close, with a larger difference on the red-wing side. As mentioned in Section.~\ref{sec:limitations}, it could mainly be the result of not using an accurate H$_2$O spectrum (here, it is for pure water) and inaccurate or missing grain-scattering effects. Most of the other species are also expected to be affected by these effects.

The main point here is to show how the synthetic spectra can help astronomers to localise features and test their model predictions of certain species. It can also facilitate our understanding of what kind of grain population we are probing, as we show that the dust temperature strongly affects the ice chemistry. As SynthIceSpec is flexible and easy to use, it can help us to understand the material distribution along the LOS; with future updates allowing us to model the grain growth, we expect to help modelers to benchmark their simulations and add constraints to both the chemistry and physics at play.

\subsection{Predicting new species in ices}\label{detection_ch3cn}
    
\citet{Taillard_2025} presented an extensive example of how to use SynthIceSpec to predict the column densities needed for new detections. In the present paper, we show how it can be applied to another molecule, CH$_3$CN, but this time by considering the impact of stellar emission on the band identification.

\subsubsection{Estimating CH$_3$CN detectability}

\begin{figure}[!h]
    \centering
    \includegraphics[width=0.7\linewidth]{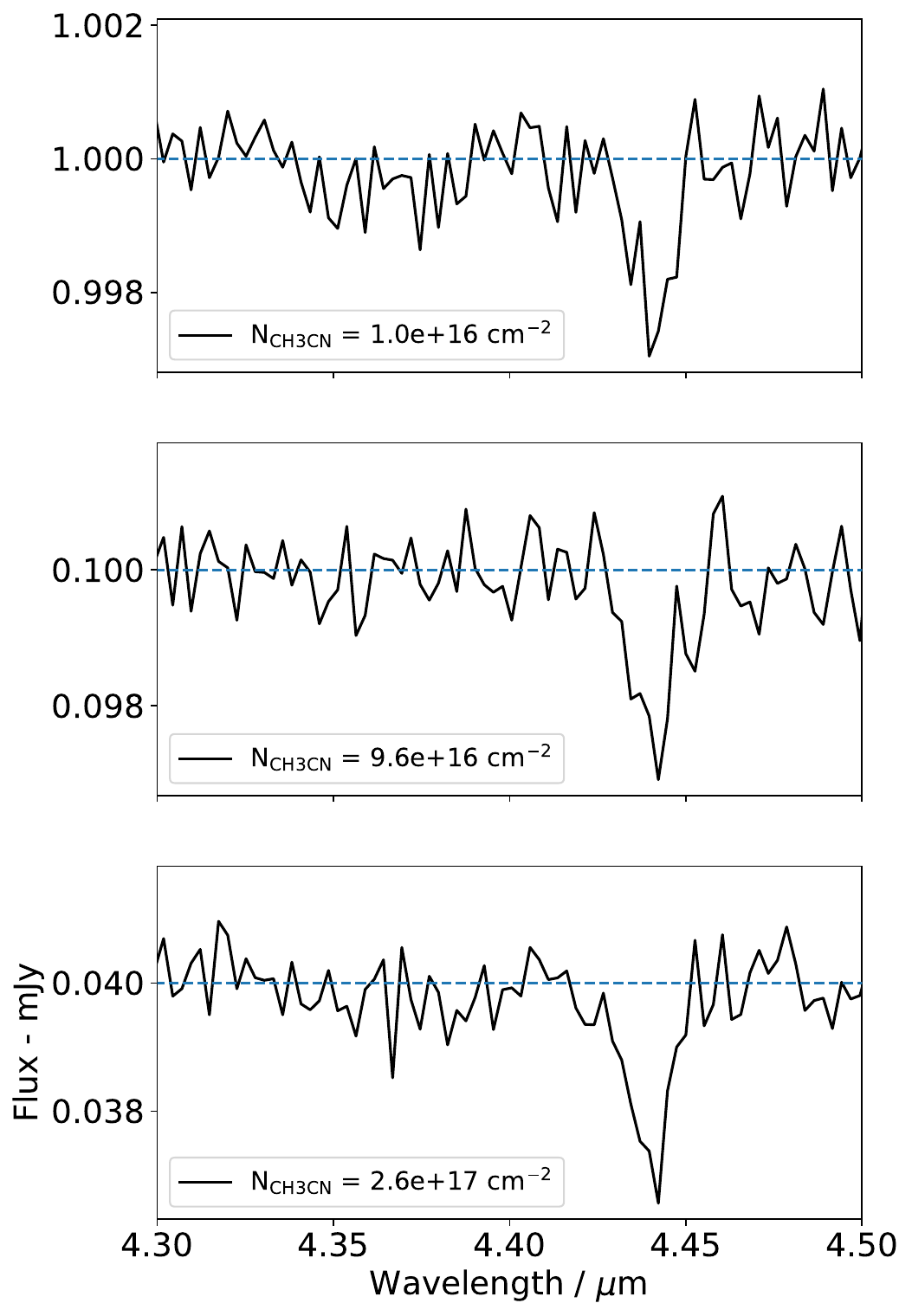}
    \caption{Synthetic spectra using NIRSpec/G395H instrument mode. The detection threshold for the 4.45 $\mu$m feature of CH$_3$CN for three different flat continua (from top to bottom: 1, 0.1, and 0.04 mJy) used to model different targets with decreasing brightnesses is shown. A synthetic noise extracted from previous observations \citep{Mcclure_2023} was added to the continuum. The black spectrum is the column density of CH$_3$CN that was increased iteratively to achieve a 5$\sigma$ detection with respect to the flat continuum (dashed blue line).}
    \label{fig:Figure_threshold}
\end{figure}

Methyl cyanide (CH$_3$CN) constitutes a key molecule for solid-phase detection as it is one of the most abundant COMs detected in the gas phase in various environments such as the Galactic Center \citep{Solomon1971}, cold cores \citep{Potapov_2016,Spezzano_2017}, protostars \citep{Rosero_2013,Jaber_2014,Andron_2018}, hot corinos \citep{Remijan_2004,Martin-Domenech_2021}, and protoplanetary discs \citep{Oberg_2015,Johnston_2015}. 
CH$_3$CN presents a strong oscillator -CN group that is observable with IR spectroscopy and with the band associated with this functional group located around 4.44 $\mu$m. We used the experimental IR spectrum measured by \citet{Rachid2022}, from which we derived the Gaussian parameters of the molecule (listed in Appendix. \ref{appendix-data}) to measure its detectability with the JWST. 
In cold cores, \citet{Mcclure_2023} reported upper-limits for CH$_3$CN towards two background stars (NIR38 and J110621) in Chameleon I using the NIRSpec FS/G395H filter. The upper limits derived are about 1.4 $\times$ 10$^{17}$ (2.03\% for H$_2$O) and 1.9 $\times$ 10$^{17}$ cm$^{-2}$ (1.37\% for H$_2$O) for NIR38 and J110621, respectively. 
Two upper limits and three tentative detections of the species were reported towards five protostars in \citet{Nazari_2024}, the authors of which used JWST/NIRSpec. However, they tentatively identified the species, as it is likely mixed with other CN-bearing molecules and species (C$_2$H$_5$CN and NO$_2$ were possible fits according to the them). The detection of CH$_3$CN was challenged by the thermal processing and mixing with other species (such as H$_2$O, CO, and CO$_2$) that affect the main vibrational mode. The CN-group oscillation was also not wide enough to explain the full observed absorption at these wavelengths. The authors derived possible column densities ranging from 2.0 to 4.3 $\times$ 10$^{17}$ cm$^{-2}$, with around 50\% uncertainties. 

As a case study, we considered the detectability of CH$_3$CN in a dense cloud environment where the thermal processing of ices has not yet started. Its formation route is difficult to trace; in the gas phase it is formed through the radiative recombination of HCN and CH$_3^+$  \citep{Giani_2023_CH3CN}, and on the surface it is predicted to be the product of successive hydrogenations of C$_2$N \citep{Garrod_2008}. However, which of the mechanisms is dominant remains a matter of debate. 
According to certain models \citep{Garrod_2022}, it is expected to form in large quantities during the cold-core phase before the collapse of the cloud, leading to the release of the molecule into the gas phase; it is subsequently destroyed. We assume that CH$_3$CN is largely produced in ices in cold-core environments. Dense cloud cases, which are often observed thanks to background stars, also allow us to consider a main limitation to COM detectability: the photosphere emission lines that can hinder its identification. In this regard, we modelled three different fiducial fluxes (1, 0.1, and 0.04 mJy) to estimate the 5$\sigma$ detection thresholds for different targets with decreasing brightnesses. For each, we iteratively increased the column density and measured the signal-to-noise ratio (S/R). The ice composition we input is the same as the one reported in NIR38 from \citet{Mcclure_2023}, Table. 2.
Since the main vibrational mode is at $\sim$ 4.45 $\mu$m, the closest feature is $^{13}$CO$_2$ at 4.38$\mu$m. The latter is rather narrow and does not overlap in our case. 
We derived the detection thresholds of the molecule with the NIRSpec/395H instrument parameters; this is the highest spectral resolution mode available with NIRSpec. 
The estimated 5$\sigma$ detection thresholds are 1.0 $\times$ 10$^{16}$, 9.6 $\times$ 10$^{16}$, and 2.6 $\times$ 10$^{17}$ cm$^{-2}$ for the 1, 0.1, and 0.04 mJy fluxes, respectively. In Figure.~\ref{fig:Figure_threshold}, we show the different thresholds for each flux, with a flat continuum shown by dashed blue lines and the spectrum containing the column density derived for each threshold and the synthetic noise shown in black. 
In \citet{Mcclure_2023}, the authors derived two 1$\sigma$ upper limits by computing the column density, N = RMS $\times$ FWHM/A, with the FWHM and band strength, A, derived from the same spectrum that we used \citep{Rachid2022}. The authors found values of 1.4 $\times$ 10$^{17}$ and 1.9 $\times$ 10$^{17}$ cm$^{-2}$ towards NIR38 and J110621, respectively. The computation was done before removing the stellar photosphere that may contribute to the RMS, making it hard to consider as a solid value. 
Our thresholds are lower than the values measured towards protostars in \citet{Nazari_2024}, but in this study, the derived continuum presents high uncertainties; this impacts the optical depth and the derived column densities, while the contribution of other species is also uncertain. \citet{Nazari_2024} found that it is also uncertain what kind of matrix the CH$_3$CN is supposed to be mixed with, and the analysis is limited by the laboratory data available.

It is important to take into consideration the contribution of the CH$_3$CN feature in the 4.45 $\mu$m region. The derived thresholds should be taken as lower limits, as the code is still missing key features that could hinder the detection (other CN-carriers, mixing, etc.). 

\subsubsection{Considering the stellar photosphere}

\begin{figure}[!h]
    \centering
    \includegraphics[width=0.7\linewidth]{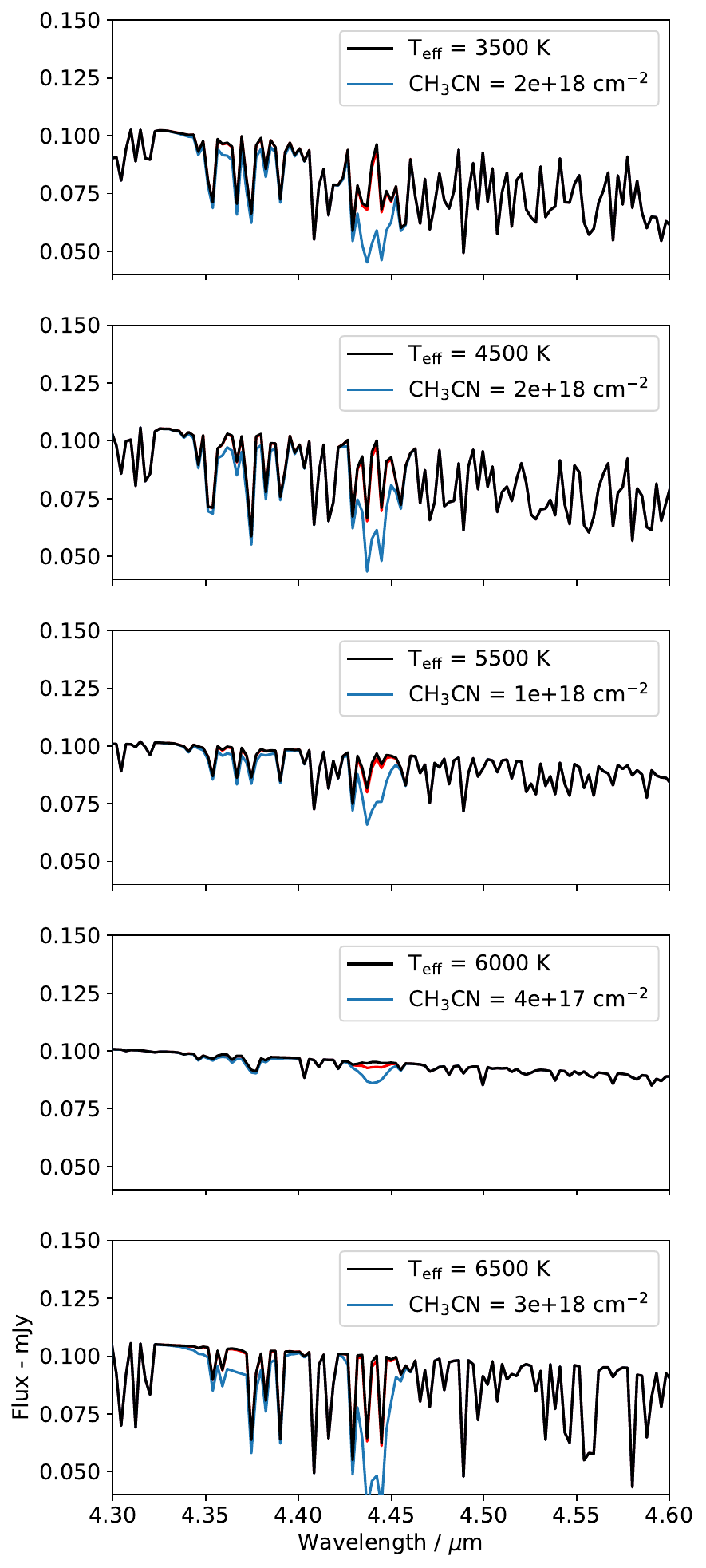}
    \caption{Synthetic ice spectra to which a photostellar spectrum at five different effective temperatures was added, containing the column densities listed in Table. 2 of \citet{Mcclure_2023} in black (with N$\rm_{CH_3CN}$ = 0), the previously estimated 5$\sigma$ threshold in red (computed without including the photosphere contribution, see Fig.~\ref{fig:Figure_threshold}), and an example of the column density needed to proceed to the identification of the 4.45 $\mu$m CH$_3$CN feature in blue. Each spectrum presents its own effective temperature for the stellar model used (from top to bottom: T$_{eff}$ = 3500, 4500, 5500, 6000, and 6500 K), to which we applied an extinction law for different A$_V$ (25, 33, 40, 42, and 45 mag respectively).}
    \label{fig:Figure_photosphere}
\end{figure}

Since photospheric lines from the background star providing the continuum are likely to be present, we checked whether our 5$\sigma$ detection thresholds would still hold with the addition of a stellar spectrum contribution. In this instance, we picked five different surface temperatures with the PHOENIX model and added them to the previous spectra that only considered ice absorption and noise on a flat, constant continuum. 
In Fig.~\ref{fig:Figure_photosphere}, we plot five different spectra composed of the ice absorption and noise, to which we added a photosphere model. The stellar spectra have an increasing effective temperature (T$_{eff}$ = 3500, 4500, 5500, 6000, and 6500 K), to which an extinction was applied using the formula in Eq.~\ref{eq_extinction_curve} with a different A$_V$ for each (25, 33, 40, 42, and 45 mag, respectively), reaching an average of 0.1 mJy. 
For each spectrum, the contribution of the 5$\sigma$ threshold for a previously computed flux of 0.1 mJy (9.6 $\times$ 10$^{16}$ cm$^{-2}$) is shown in red. We then checked whether or not the stellar type is suitable for detection. In most cases, the photosphere features are stronger than the CH$_3$CN absorption by a factor between ten and 100, leaving it impossible to even identify the species. In blue, we plot an example of the column density required in order to detect the species, leading in all cases (except for T$_{eff}$ = 6000 K) to very high values that are unrealistic compared to model predictions and previously set upper limits from observations. It seems that the T$_{eff}$ = 6000 K case (fourth panel) is the most favourable for an identification and possible detection. The photospheric lines are thus a strong obstacle to the detection of this feature in particular. 

Overall, the detection of CH$_3$CN is expected to be difficult in dense cloud environments, mainly because of the stellar spectrum contributions needed as background sources. Modelling and subtracting the photosphere contribution is thus essential in order to properly measure CH$_3$CN contribution at 4.45 $\mu$m. 
The model we used for photosphere absorption lines could have limitations arising from assumptions used to produce them in the first place (e.g. local thermodynamical equilibrium versus non-local
thermodynamical equilibrium). 
As new stellar photosphere models are optimised with new data \citep[e.g.][]{Meszaros_2024}, and a stellar spectral library compiled with JWST data will soon be available, we will update SynthIceSpec accordingly to facilitate the search for new species never before seen in ices. As long as one can correct for the photospheric lines, they are not a limitation to the detection. 

\section{Conclusions}

In this paper, we present a synthetic ice spectra generator, SynthIceSpec, which was designed to facilitate the interpretation of JWST data and the prediction of new species detection. The code is based on the approximation that ice's vibrational mode can be represented by a Gaussian or decomposed in a sum of Gaussians, described by a band strength, a bandwidth, and a position on the IR spectrum. These parameters were retrieved from laboratory spectra.
The code produces spectra corresponding to the wavelengths and spectral resolutions of the various JWST spectroscopic instruments. 
In order to simulate more realistic observations, a contribution from the silicate absorption and from different astrophysical continuum sources (stellar photospheric spectra, black bodies, protoplanetary disc continuum) can be added.
The model takes the column densities of molecules defined by the user or read as output from a chemical model (such as Nautilus) as input. The list of Gaussian parameters, IceSpecData, is available for the pure spectra of many solid species, and is expanding with the addition of certain ice mixtures (which have an impact on the spectral parameters). The code is public and available online; it will be updated regularly to introduce missing parameters and phenomena such as grain growth, mixing, new continua, and species.
It can be used for different purposes, such as the post-processing of chemical-model predictions, identification of molecules, comparison with observations, testing of the detectability of new species, and so on. 
In this paper, we show two different use cases that highlight the following astrochemical implications:

\begin{itemize}
    \item We ran gas-grain models to understand the impact of dust temperature on CO$_2$ formation and compare it with past observations. We tested three different grain temperatures (6.6, 10.3, and 13.2 K), where the two extrema either produced very little CO$_2$ or too much. None of the model can simultaneously reproduce the three species that we studied (H$_2$O, CO$_2,$ and CH$_3$OH).
    
    \item The intermediate temperature of 10.3 K is able to reproduce the CO$_2$ observed column density very
well. The dust temperature strongly impacts its formation in chemical models, where a shift of only a few kelvin could drift towards a very different ice composition that is easily illustrated with synthetic spectra, completely changing the absorption profiles. New constraints in models (grain size distribution, sticking coefficient, binding energies, etc.) must be added to fully encapsulate CO$_2$ formation.
    
    \item Solid CO$_2$ observations could add constraints to the dust temperature. Coupled with the grain growth impacting its absorption-band shape at 4.2 $\mu$m, it can provide important information regarding the grain size distribution being probed along the LOS. Spatial ice mapping with the JWST could unveil the dust's physical properties as well as the ice composition and role of CO$_2$. 

    \item We checked the detectability of the methyl-cyanide (CH$_3$CN) feature at 4.45 $\mu$m with the JWST in cold-core environments, testing three different flat fluxes and adding fiducial noise. The estimated 5$\sigma$ detection thresholds are 1.0 $\times$ 10$^{16}$, 9.6 $\times$ 10$^{16}$, and 2.6 $\times$ 10$^{17}$ cm$^{-2}$ for the 1, 0.1, and 0.04 mJy flux, respectively. These column densities are lower than previous upper limits obtained in cold cores and YSOs, where the latter were obtained considering strong observational barriers (e.g. photospheric lines, complicated continuum, etc.). If formed efficiently, CH$_3$CN could be observed as this main feature is not expected to be overlapped by other species in cold environments ($>$ 0.1 \% with regard to H$_2$O in the absence of any overlap or continuum contribution).

    \item The stellar spectrum and, more precisely, photospheric lines strongly hinder the detection of the feature depending, for example, on the effective temperature (e.g. $>$ 14.4\% with regard to H$_2$O is needed to assert a 5$\sigma$ detection). We expect that new photospheric library and models will provide better tools to model the photosphere and will be able to correct the JWST spectra efficiently to identify weaker features such as CH$_3$CN.
    
\end{itemize}

We hope to provide an easy-to-use tool for the community that can be used to constrain the feasibility for proposals and interpret JWST data. SynthIceSpec can be easily used in the JWST exposure-time calculator to help astrochemists prepare their observations. New updates will be published as soon as grain growth models, stellar models, extinction laws, and laboratory spectra become available. We encourage spectroscopists, observers, and modelers to participate to the development of this open-source tool.

\section*{Data availability}
SynthIceSpec source code is available on Zenodo:
\url{https://zenodo.org/records/18460473}.

\begin{acknowledgements} 
This project has received funding from the European Research Council (ERC) under the European Union’s Horizon Europe research and innovation programme ERC-AdG-2022 (GA No. 101096293). AT, PG, JAN, ED, AC and VW acknowledge the thematic Action “Physique et Chimie du Milieu Interstellaire” (PCMI) of INSU Programme National “Astro”, with contributions from CNRS Physique \& CNRS Chimie, CEA, and CNES.

\end{acknowledgements}

\bibliography{biblio}
\bibliographystyle{aa}

\appendix

\onecolumn

\section{Spectral data used in SynthIceSpec}\label{appendix-data}

\begin{table*}[!h]
\centering
\caption{Gaussian parameters used in SynthIceSpec}
\begin{tabular}{lcccccc}
\hline
\hline
Molecule & Position   & Position & FWHM        & Gaussian intensity  & Mode      & Reference       \\ 
         & (cm$^{-1}$)& ($\mu$m) & (cm$^{-1}$) &   (cm/molecule)         &                   \\
\hline
  H$_2$O & 3239.8 & 3.1 & 222.9 & 1.2E-16 & O-H stretch &  \text{\citet{Oberg_2007}} \\
  H$_2$O & 3398.6 & 2.9 & 207.9 & 6.9E-17 &  O-H stretch & \text{\citet{Oberg_2007}} \\ 
  H$_2$O & 3408.4 & 2.9 & 355.3 & 2.4E-17 &  O-H stretch & \text{\citet{Oberg_2007}} \\
  H$_2$O & 3667.0 & 2.7 & 20.9 & 3.3E-19  & Dangling O-H   & \text{\citet{Oberg_2007}} \\
  H$_2$O & 3688.0 & 2.7 & 23.5 & 3.3E-19 & Dangling O-H  & \text{\citet{Oberg_2007}} \\
  H$_2$O & 1660.0 & 6.0 & 160.0 & 1.0E-17 & O-H bending & \text{\citet{Oberg_2007}} \\
  H$_2$O & 765.0 & 13.1 & 246.0 & 1.2E-18 & Libration & \text{\citet{Oberg_2007}}\\
  CO    & 2140.0 & 4.7 & 5.0 & 1.1E-17 & C-O stretch & \text{\citet{1995A&A...296..810G}}\\
  CO$_2$ & 2343.0& 4.2 & 18.0 & 7.6E-17 & C-O stretch &  \text{\citet{1995A&A...296..810G}}\\
  CO$_2$ & 660.0 & 15.1 & 18.0 & 1.2E-17 & C-O bending & \text{\citet{1995A&A...296..810G}} \\
  CH$_4$ & 3010.0 & 3.3 & 7.0 & 7.0E-18 & C-H stretch & \text{\citet{1993ApJS...86..713H}}\\
  CH$_4$ & 1302.0 & 7.6 & 8.0 & 7.0E-18 & C-H bending & \text{\citet{1993ApJS...86..713H}}\\
  CH$_3$OH & 3250.0 & 3.1 & 235.0 & 1.1E-16 & O-H stretch  & \text{\citet{2023MNRAS.525.2690C}}\\
  CH$_3$OH & 2982.0 & 3.3 & 100.0 & 2.1E-17 &  C-H stretch & \text{\citet{2023MNRAS.525.2690C}}\\
  CH$_3$OH & 2828.0 & 3.5 & 30.0 & 8.0E-18 & C-H stretch & \text{\citet{2023MNRAS.525.2690C}}\\
  CH$_3$OH & 1460.0 & 6.8 & 90.0 & 1.0E-17 & C-H bending & \text{\citet{2023MNRAS.525.2690C}}\\
  CH$_3$OH & 1130.0 & 8.8 & 34.0 & 1.4E-18 & CH$_3$ rocking & \text{\citet{2023MNRAS.525.2690C}}\\
  CH$_3$OH & 1030.0 & 9.7 & 29.0 & 1.4E-17 & C-O stretching & \text{\citet{2023MNRAS.525.2690C}}\\
  CH$_3$OH & 700.0 & 14.3 & 200.0 & 1.6E-17 & torsion & \text{\citet{2023MNRAS.525.2690C}}\\
  NH$_3$ & 3375.0 & 2.9 & 45.0 & 2.3E-17 & N-H stretch & \text{\citet{2013MNRAS.428.3262N}}\\
  NH$_3$ & 1630.0 & 6.1 & 60.0 & 5.0E-18 & N-H bending & \text{\citet{2013MNRAS.428.3262N}}\\
  NH$_3$ & 1070.0 & 9.3 & 70.0 & 1.7E-17 & Umbrella & \text{\citet{2013MNRAS.428.3262N}}\\
  HCOOH & 1714.0 & 5.8 & 45.0 & 6.7E-17 & C=O stretch & \text{\citet{2015MNRAS.451.2145B}}\\
  HCOOH & 1650.0 & 6.0 & 69.0 & 6.7E-17 & Combination & \text{\citet{2015MNRAS.451.2145B}}\\
  HCOOH & 1387.0 & 7.2 & 37.0 & 3.6E-17 & C-H bending & \text{\citet{2015MNRAS.451.2145B}}\\
  HCOOH & 1211.0 & 8.2 & 64.0 & 2.8E-18 & C-O stretch & \text{\citet{2015MNRAS.451.2145B}}\\
  HNCO & 2240.0 & 4.6 & 56.0 & 7.8E-17 & C-O stretch & \text{\citet{2016MNRAS.460.4297F}}\\
  HCN & 2092.5 & 4.7 & 20.0 & 1.1E-17 & C-N stretch & \text{\citet{2022MNRAS.509.3515G}}\\
  H$_2$CO & 2992.0 & 3.3 & 17.0 & 1.2E-18 & Combination & \text{\citet{2015MNRAS.451.2145B}}\\
  H$_2$CO & 2881.0 & 3.4 & 21.0 & 2.7E-18 & C-H$_2$ stretch & \text{\citet{2015MNRAS.451.2145B}}\\
  H$_2$CO & 2819.0 & 3.5  & 28.0 & 3.7E-18 & C-H$_2$ stretch & \text{\citet{2015MNRAS.451.2145B}}\\
  H$_2$CO & 1720.0 & 5.8 & 20.0 & 9.6E-18 & C=O stretch & \text{\citet{2015MNRAS.451.2145B}}\\
  H$_2$CO & 1494.0 & 6.7 & 15.0 & 1.0E-18 & C-H$_2$ scissoring &  \text{\citet{2015MNRAS.451.2145B}} \\
  NH$_2$CH$_2$OH & 1610.0 & 6.2 & 45.0 & 4.6E-18 & N-H bending & \text{\citet{2009ApJ...707.1524B}} \\
  NH$_2$CH$_2$OH & 1470.0 & 6.8 & 40.0 & 7.0E-18 & O-H bending & \text{\citet{2009ApJ...707.1524B}}\\
  NH$_2$CH$_2$OH & 1390.0 & 7.2  & 15.0 & 2.5E-18 & torsion & \text{\citet{2009ApJ...707.1524B}}\\
  NH$_2$CH$_2$OH & 1007.0 & 9.9 & 145.0 & 3.5E-17 & C-N stretch & \text{\citet{2009ApJ...707.1524B}}\\
  HDO & 2457.0 & 4.1 & 120.7 & 4.3E-17 & O-D stretch & \text{\citet{2012A&A...538A..57A}}\\
  $^{13}$CO & 2092.0 & 4.8 & 1.5 & 1.1E-17 & $^{13}$C-O stretch & \text{\citet{2002ApJ...577..271B}} \\
  CH$_3$CHO & 1721.0 & 5.8 & 17.3 & 1.06E-17 & C=O stretching & \text{\citet{Terwisscha-van-Scheltinga_2018}} \\
  CH$_3$CHO & 1347.0 & 7.4 & 13.1 & 2.7E-18 & C-H bending & \text{\citet{Terwisscha-van-Scheltinga_2018}} \\
  CH$_3$CHO & 1122.0 & 8.9 & 12.4 & 1.7E-18 & CH$_3$ rocking & \text{\citet{Terwisscha-van-Scheltinga_2018}} \\
  H$_2$S & 2545.0 & 3.9 & 45.0 & 3.0E-17 & H-S stretching &  \text{\citet{2022ApJ...931L...4Y}}\\
  OCS & 514.0 & 19.4 & 18.2 & 1.8E-18 & C-S stretch & \text{\citet{1993ApJS...86..713H}} \\
  OCS & 2025.0 & 4.9 & 22.5 & 1.5E-16 & C-S stretch & \text{\citet{1993ApJS...86..713H}} \\
  CS$_2$ & 1502.0 & 6.6 &  23.0 & 1.06E-16   &  C-S stretch & \text{\citet{Taillard_2025}} \\
  CS$_2$ & 2145.0 & 4.6 &  32.8 &  2.09E-18   &  C-S stretch & \text{\citet{Taillard_2025}} \\
  S$_8$  & 471.0 & 21.2 &  14.8 & 1.00E-19 & S-S stretch &  \text{\citet{Taillard_2025}} \\
  CH$_3$SH    & 2535 & 3.9 & 40 &  5.41E-18 & S-H stretch & \text{\citet{Hudson_2018}}  \\
  CH$_3$CH$_2$SH & 2528 & 3.9 & 40 &  4.2E-18  & S-H stretch &  \text{\citet{Hudson_2018}} \\ 
  SO$_2$ & 1339.0 & 7.4 & 9.8 & 1.5E-17 & S-O stretch &  \text{\citet{2008P&SS...56.1300G}} \\
  SO$_2$ & 1149.0 & 8.7 & 8.7 & 2.2E-18 & S-O stretch & \text{\citet{2008P&SS...56.1300G}}\\
  SO & 1130.0 & 8.8 & 4.3 & 2.2E-18 & S-O stretch & \text{\citet{2008P&SS...56.1300G}}\\

\hline\end{tabular}
\tablefoot{Spectral data extracted from the Gaussian fitting of pure ice species, we list the molecule vibration mode, its wave number, position in microns, band width, band strength and reference associated. }
\label{tab:gaussparam1}
\end{table*}

\begin{table*}[!h]
\centering
\caption{Gaussian parameters used in SynthIceSpec}
\begin{tabular}{lcccccc}
\hline
\hline
Molecule & Position   & Position & FWHM        & Gaussian intensity  & Mode      & Reference       \\ 
         & (cm$^{-1}$)& ($\mu$m) & (cm$^{-1}$) &   (cm/molecule)         &                   \\
\hline
  OCN$^-$ & 2175.0 & 4.6 & 20.0 & 1.3E-16 & C-N stretch & \text{\citet{Mcclure_2023}}\\
  NH$_4^+$ & 1459.0 & 6.8 & 64.0 & 4.4E-17 & N-H stretch & \text{\citet{2003A&A...398.1049S}}\\
  CH$_3$CN & 3002 & 3.3  & 18.1 & 1.50E-18 & CH$_3$ stretch &\text{\citet{Rachid2022}} \\
  CH$_3$CN & 2942 & 3.4  & 9.4  & 0.53E-18 & CH$_3$ stretch & \text{\citet{Rachid2022}} \\
  CH$_3$CN & 2289 & 4.3  & 16.4 & 0.60E-18 & Combination & \text{\citet{Rachid2022}}  \\
  CH$_3$CN & 2252 & 4.4  & 7.0  & 1.90E-18 & C-N stretch & \text{\citet{Rachid2022}} \\
  CH$_3$CN & 1449 & 6.9  & 18.8 & 2.90E-18 & Combination & \text{\citet{Rachid2022}}  \\
  CH$_3$CN & 1411 & 7.1  & 14.1 & 1.90E-18 & CH$_3$ deformation & \text{\citet{Rachid2022}} \\
  CH$_3$CN & 1374 & 7.3  & 16.4 & 1.20E-18 & CH$_3$ deformation & \text{\citet{Rachid2022}} \\
  CH$_3$CN & 1042 & 9.6  & 18.8 & 1.60E-18 & CH$_3$ rock & \text{\citet{Rachid2022}} \\
  CH$_3$CN & 920  & 10.9  & 7.0  & 0.35E-18 & C-C stretch & \text{\citet{Rachid2022}} \\
  HCOOCH$_3$  &  1722.0 &  5.8 &  28.2  &  4.9E-17 & C=O stretch & \citet{Terwisscha2021} \\
  HCOOCH$_3$  &  1215.0 & 8.2  &  23.5  &  2.9E-17 & C-O stretch & \citet{Terwisscha2021}\\ 
  HCOOCH$_3$  &  1163.0 & 8.5  &  25.8  &  1.9E-17 & CH$_3$ rock & \citet{Terwisscha2021} \\
  HCOOCH$_3$  &  910.7  & 10.9 &  16.4  &  4.8E-18 & O-CH$_3$ stretch & \citet{Terwisscha2021} \\ 
  HCOOCH$_3$  &  768.3  & 13.0 &  7     &  1.2E-18 & deformation & \citet{Terwisscha2021} \\
\hline\end{tabular}
\tablefoot{Follow-up to Table.~\ref{tab:gaussparam1}.}
\label{tab:gaussparam2}
\end{table*}

\twocolumn

\section{SynthIceSpec step-by-step}\label{appendix-decomposition}

In this Appendix, we show how each component (ices, silicates, noise, stellar spectrum and extinction) are added one by one, plotted as the flux (mJy) as a function of the wavelength ($\mu$m). In the first plot,  only the ice absorption is shown. In the second plot, we add the noise and silicates to the ice absorption. Finally, in the third plot, we add the stellar spectrum and the extinction formalism derived in Section.~\ref{sec:extinction_law}.

\begin{figure}[!h]
    \centering
    \includegraphics[width=0.9\linewidth]{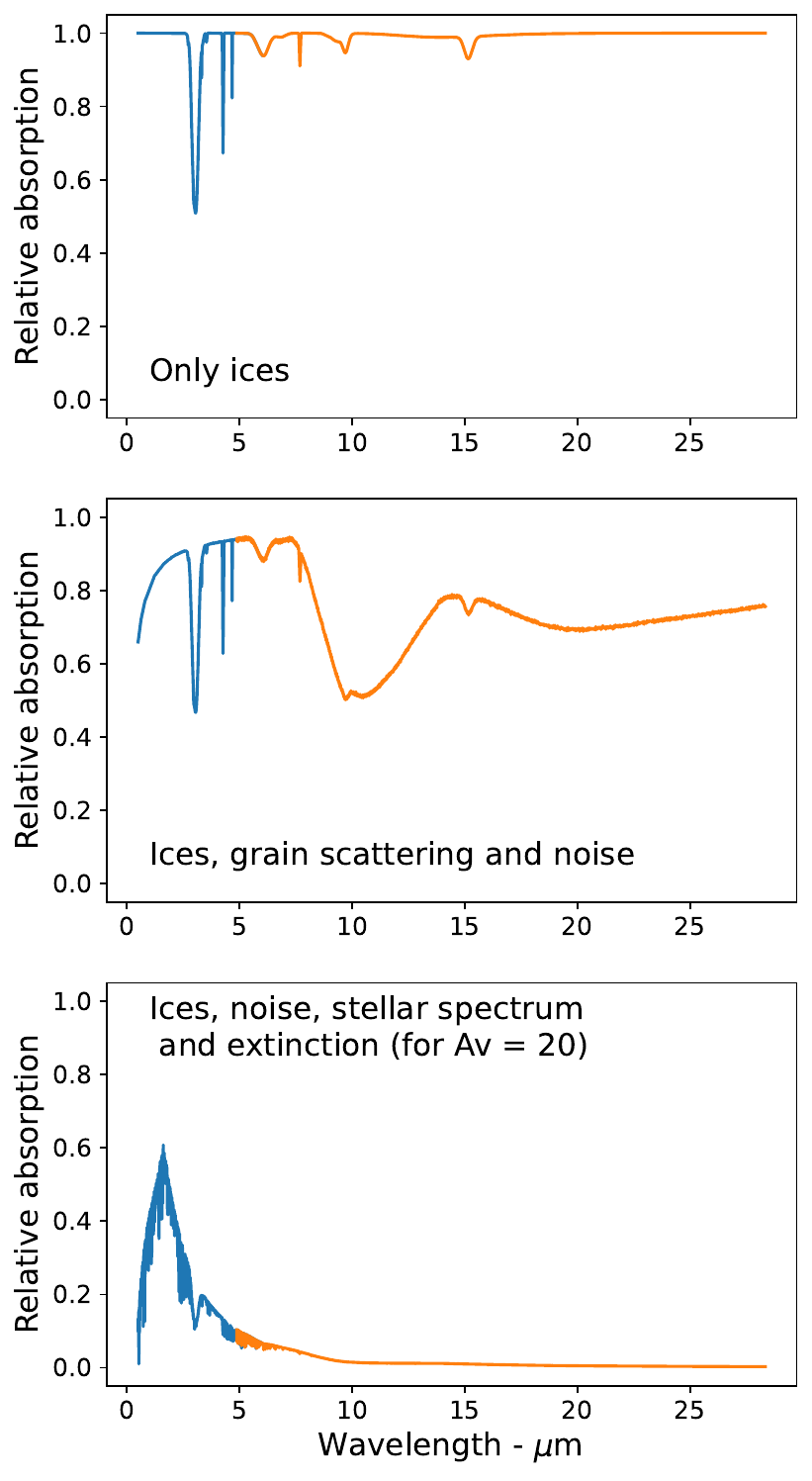}
    \caption{Flux (mJy) as a function of the wavelength ($\mu$m), from top to bottom: the first panel only shows ice absorption, the second panel is the ice absorption to which is applied the grain scattering and instrumental noise, finally the third show a stellar spectrum and extinction law applied to ice absorption.}
    \label{fig:decomposition}
\end{figure}

\section{Chemical post-processing examples}\label{appendixB}

\begin{figure}
    \centering    
    \includegraphics[width=0.99\linewidth]{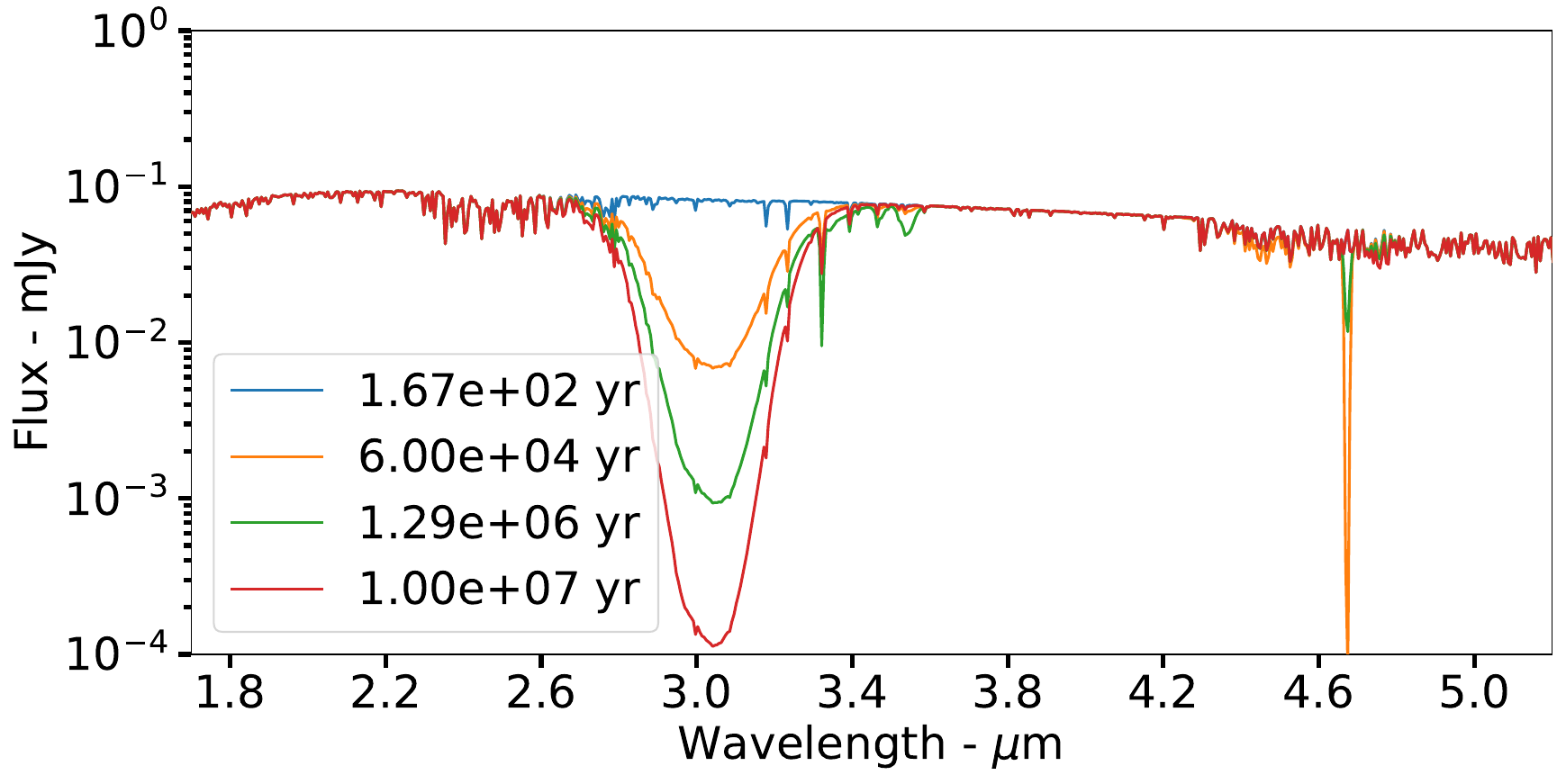}
    \caption{Nautilus predictions of the core L429-C from \citet{Taillard_2023} with the NIRSpec instrument, showing the different ice composition as predicted by the model for four different times. A K7 stellar type is applied to the ice absorption features and an extinction for an A$_V$ of 24.2 is applied on the continuum. }
    \label{fig:nautilus_prediction_time}
\end{figure}

Since gas-grain models predicts abundances as a function of time, it is possible to follow how the ice spectrum will look like for each time-step. The abundances themselves are converted to column densities by reading the input parameter in the model. We show in Fig.~\ref{fig:nautilus_prediction_time} an example of the evolution of ice composition at four different times, with the flux as a function, where we applied a K7 stellar type spectrum on the predicted abundances. The parameters of the model used here are: A$_V$ = 24.2, T$_{dust}$ = 14.5, N$\rm_{H_2}$ = 2.1 $\times$ 10$^{22}$ cm$^{-3}$, $\zeta_{CR}$ = 3 $\times$ 10$^{-17}$ s$^{-1}$. 
This simple use can help to tract down which ice species might dominate the ice composition as function of time and physical parameters. It can be used for different initial conditions to even allow to track the evolution from a diffuse to dense cloud. 

\section{Fixed dust grain temperature models outputs}\label{dust-models}

\begin{figure}
    \centering
    \includegraphics[width=0.99\linewidth]{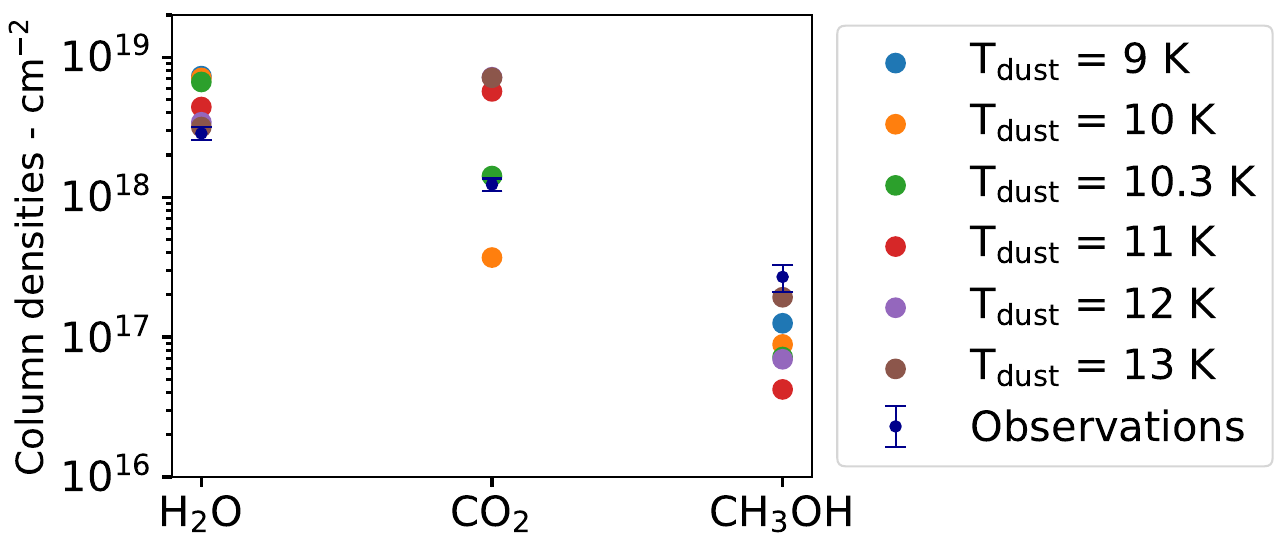}
    \caption{Comparison between the different fixed dust temperature models where the modelled column densities for H$_2$O, CO$_2$ and CH$_3$OH is plotted as a coloured dot for each model and the observed column densities from \citet{Boogert_2011} are plotted in dark blue with their uncertainties. }
    \label{fig:dust-models}
\end{figure}

In this Appendix, we present the outputs of the static models, mentioned in \ref{CO-study}, where we fixed the dust temperature in a range between 9 to 13 K. For each model, the abundance used to compare with the observed column densities derived in \citet{Boogert_2011} of the background star J18170957 were obtained at t = 1 $\times$ 10$^5$ years. In Figure.~\ref{fig:dust-models}, we compare the observed column densities (as coloured lines) of H$_2$O, CO$_2$ and CH$_3$OH for each model (coloured dot corresponding to a fixed temperature). 

\onecolumn
\pagebreak

\end{document}